\documentclass[manuscript,screen]{acmart}

\acmJournal{TOSEM}
\acmYear{2026}
\copyrightyear{2026}
\setcopyright{none}                 % placeholder for a working draft
\usepackage{graphicx}
\usepackage{booktabs}
\usepackage{tabularx}
\usepackage{longtable}
\usepackage{pifont}
\usepackage{newunicodechar}
\newunicodechar{×}{\ensuremath{\times}}
\newunicodechar{→}{\ensuremath{\rightarrow}}
\newunicodechar{≥}{\ensuremath{\ge}}
\newunicodechar{≤}{\ensuremath{\le}}
\newunicodechar{·}{\ensuremath{\cdot}}
\newunicodechar{§}{\S}

\providecommand{\tightlist}{\setlength{\itemsep}{0pt}\setlength{\parskip}{0pt}}

\begin{document}

%\title{The Data Problem in Software Vulnerability Analysis:
%       A Survey of Artifacts, Quality, and Consumption}

\title{The Data Problem in Software Vulnerability Analysis:
       Artifacts, Quality, and Consumption}

\author{Yu Nong}
\affiliation{%
  \institution{Oakland University}
  \city{Rochester Hills}\state{Michigan}\country{USA}}
\email{yunong@oakland.edu}

\author{Yao Du}
\affiliation{%
  \institution{Macau University of Science and Technology}
  \city{Macau}\country{Macau SAR}}
\email{yaodu19@gmail.com}

\author{Tianxiang Xu}
\affiliation{%
  \institution{Kent State University}
  \city{Kent}\state{Ohio}\country{USA}}
\email{txu8@kent.edu}

\author{Haipeng Cai}
\affiliation{%
  \institution{University at Buffalo, SUNY}
  \city{Buffalo}\state{New York}\country{USA}}
\email{haipengc@buffalo.edu}

\renewcommand{\shortauthors}{Nong et al.}

\begin{abstract}
Learning- and LLM-based software vulnerability analysis is only as trustworthy as the data it is
trained and evaluated on, yet that data is rarely examined as a first-class object. We investigate the
\emph{data} behind vulnerability analysis through a dataset-centric taxonomy that separates what an
artifact is (code, metadata, patches, tests/PoCs, reasoning, traces), how good it is (realism, label evidence, scale, diversity, leakage, availability), and what it is used for. From a systematically assembled corpus of 1{,}522 papers covering 2016--2026 plus foundational earlier work we deep-code a tiered set of 111 anchor papers, backing every affirmative rubric-graded value with a verbatim span, and we report, per attribute, both how much it has been
\emph{studied} and how well datasets \emph{achieve} it. The results trace an \emph{evidence ladder}:
executable artifacts are the only major type where 15 of the 24 datasets are both graded real-world and carry labels that received an independent check, while code-sample datasets---the largest category in both the auto-tagged corpus and the anchor set---are the least realistic: 20 of the 41 draw their vulnerabilities from authentic projects or CVEs, but only 3 keep the sample at the unit the code is deployed in, and only 2 do both---though these are coarse component tests, and just one code-sample dataset meets the codebook's stricter full-context real-world grade. Among these, leakage goes unaddressed by 49 of the 90 datasets where it applies, more than a quarter say nothing about availability, reasoning data has arrived only recently and is mostly model-generated, and primary trace corpora remain limited to three datasets, the total after a corpus-wide screen and a full-text check of every candidate it surfaced, with further datasets releasing traces secondarily behind benchmarks and harnesses. We distill these into an
artifact$\times$quality matrix and a research agenda for evidence-rich vulnerability data.
\end{abstract}

%% CCS concepts. Regenerate the CCSXML block from ACM's CCS tool
%% (https://dl.acm.org/ccs) before final submission: the concept IDs it emits are what
%% ACM ingests, and they cannot be written by hand reliably. The \ccsdesc lines below
%% produce the printed block and compile on their own.


\maketitle

\section{Introduction}\label{sec:intro}

Software vulnerabilities remain among the most consequential and persistent threats in computing,
and the defenses against them have changed character twice within a decade. Traditional program analysis---static, dynamic, and symbolic---was increasingly complemented by \emph{data-driven} detection, in which deep neural networks learn to flag vulnerable code from large labeled corpora, and more recently by large language models (LLMs) prompted, fine-tuned, or embedded in autonomous \emph{coding
agents} that now write, patch, and review software with little human supervision. What unites these
waves is a dependence that is seldom stated outright: each is only as trustworthy as the
vulnerability \emph{data} it was trained and evaluated on.

Yet that data is rarely examined as an object in its own right. A detector reported at 95\% accuracy
on a popular corpus is trustworthy only insofar as the corpus's labels are correct, its samples are
realistic, its train and test splits are independent, and its class distribution matches deployment---assumptions that, across the literature, quietly fail. Detectors lose more than half of their
accuracy when moved to real-world code, in part because up to 68\% of train/test pairs are near
duplicates~\cite{pseedreveal}; label accuracy in widely used corpora falls below 60\%, and in one large patch-derived benchmark roughly one CVE in six carries a label that cannot be recovered reliably from the supplied static code alone~\cite{p951}; and for most disclosed
vulnerabilities no executable proof-of-concept exists to confirm them~\cite{p898}. Each of these was reported as a finding about one dataset or one technique. Read together, they describe a common condition. Dataset-centric studies of vulnerability data exist (Section~\ref{sec:related}); what we have not found is one that spans the full artifact range from metadata through patches, executable proofs, reasoning, and traces, and that reports for each quality attribute how much it has been \emph{studied} separately from how well datasets \emph{achieve} it.

Earlier vulnerability-analysis workflows usually placed a human between a model's output and any change to the software. Even when poor data led to an incorrect prediction, a developer could still review the result before acting on it. Coding agents increasingly remove that intermediate review step: they write patches, execute code, and modify repositories directly. Recent measurements show the consequences when these systems rely on weak evidence. Between 4.3\% and 6.0\% of \emph{functionally correct} agent patches are
vulnerable with no attacker present, and an adversarial suggestion appended to an issue description
drives that to 40.7\%~\cite{p768}. One in five agent trajectories contains an insecure action under ordinary
operation, with no adversary present~\cite{p766}; and the supply chain feeding these systems is
unguarded enough that poisoning a single retrieval sample compromises roughly half of what they
subsequently generate~\cite{p1035}. The benchmarks meant to catch such failures
inherit the same data problems they are supposed to detect. Weak-evidence training and evaluation data may now fail to detect or prevent unsafe \emph{actions} taken at machine speed and without review, rather than merely yielding a misleading number---its consequences are operational, not only statistical.

\paragraph{The evidence ladder.}
We argue that the failures above are not independent problems but reflect a common property of vulnerability data, which we call the \emph{evidence ladder}. The artifacts the field works with
range from those that merely \emph{assert} a vulnerability to those that \emph{demonstrate} it: a
CVE/CWE label or a function tagged ``vulnerable'' is a claim; a security patch narrows where the flaw
lives; a test or proof-of-concept exploit that runs demonstrates that the behaviour is reachable in the tested environment; reasoning adds explanatory content without extending that ordering, and execution traces are observational records that sit outside it. Section~\ref{sec:example} makes this concrete by rendering one vulnerability
as all six artifact types and asking, of each, what it lets a consumer conclude.

Stronger vulnerability evidence is generally more expensive to produce. At the strong-evidence end, an exploit that runs is both an artifact and independently checkable evidence that the behaviour it triggers is reachable. Producing such artifacts, however, requires substantially more effort than assigning metadata or extracting labeled code. The evidential strength of an artifact therefore affects what can be concluded about a vulnerability from that artifact alone, while production cost affects how much data of that type can practically be collected. As our study shows, the literature is largest at the weak-evidence end, where data is cheaper to obtain; executable work is now substantial in terms of paper count but remains far smaller as reusable datasets and is rarely used for training.

\paragraph{Scope.}
This study is about vulnerability \emph{data}, not about the techniques that consume it. We include
work that introduces, produces, evaluates, or improves vulnerability data, and work that reports a
finding about the quality of existing data. We exclude work that merely applies an existing dataset to
evaluate a technique without contributing to or reporting on the data itself, however strong that work
may be on its own terms; Section~\ref{survey-methodology} states the filter precisely. Detection,
repair, and generation techniques appear here only as \emph{consumers} whose requirements shape what
data is built. We cover the full artifact range---code, metadata, patches, tests and exploits,
reasoning, and traces---rather than a single artifact type or a single downstream task.

\paragraph{Closest prior work.}
The quality of vulnerability data is itself an active subject, and several studies have measured
one facet of it carefully. Croft et al. audit five widely used datasets against accuracy,
uniqueness, consistency, and currency, and find substantial defects in
each~\cite{p1456}; related audits report the same for label noise in deep-learning
corpora~\cite{p1420} and for dataset construction more broadly~\cite{p1878,p1879}. Others isolate
a single failure: duplication between training and test partitions~\cite{pseedreveal}, labels that
cannot be decided from the code they are attached to~\cite{p951}, and benchmarks whose scores
reward the wrong capability~\cite{p807}. A recent systematic review covers quality issues along
the generation-and-training pipeline for LLM-based detection~\cite{p636}. 
%We rely on these results throughout, and cite them as evidence rather than restating them.
%
These studies provide detailed evidence on individual quality dimensions, usually within code-sample datasets built for detection, but they do not systematically compare artifact types, quality properties, and downstream uses across vulnerability analysis. That leaves three questions open, and they only become visible when the artifact types are examined side by side: (1) \textit{which kinds of vulnerability artifact the literature concentrates on}, (2) \textit{whether quality varies systematically across those kinds}, and (3) \textit{whether the data being built matches what its consumers now need}. The gap matters more as new artifact types emerge: executable tests, reasoning
annotations, and agent traces are recent enough that audits of code-sample corpora say nothing
about them. The three questions %below are organized to 
are posed to address 
%answer 
exactly that.

\paragraph{Approach.}
Rather than organizing the literature by technique (detection vs.\ repair vs.\ generation) or by model
(GNN vs.\ LLM vs.\ agent), we %ask three questions  
examine three aspects 
about the data itself: what vulnerability data
exists (its \emph{artifacts}), how good it is (its \emph{quality}), and what it is used for (its
\emph{consumption}). %To answer them 
To that end, 
we construct a systematically assembled corpus of 1{,}522 papers covering 2016--2026 plus foundational earlier work, alongside the foundational earlier datasets the area still rests on; derive a taxonomy of
three aspects and six quality attributes (Section~\ref{sec:tax}); and deep-code a tiered shortlist of
111 anchor papers against it (Section~\ref{survey-methodology}). Deep coding is evidence-bound: no affirmative rubric-graded value is recorded without a verbatim span from the source, and absence values are checked by targeted full-text scans, yielding 729 quoted spans across the 111 papers. Every record was verified by the authors: 80 field by field against the source, and 31 that were first screened for conflicts, with every flag adjudicated independently by three authors at 96\% mean pairwise agreement, and then checked field by field as well. That figure is agreement on the flagged questions, not a
reliability coefficient for applying the taxonomy from scratch; Section~\ref{threats-to-validity}
reports both procedures and what each measures.

\paragraph{Findings.}
The pattern is consistent across attributes. The coded quality profile varies with demonstrative strength, most clearly at the ends of the ladder, although part of that association follows from our operational definitions (Section~\ref{survey-result-data-quality-rq2-the-core}): executable
artifacts (tests, PoCs, exploits) are the only kind where a successful run gives direct evidence that
the flaw is reachable---16 of 24 are authentically real and 23 of 24 carry labels that received an independent check---whereas code-sample datasets---the largest category in both the auto-tagged corpus and the anchor set, are the weakest on realism, of 41 such datasets we
deep-code, \emph{one} uses authentically real-world data, and only because its samples are whole
smart contracts rather than functions sliced from a project. Two recurring omissions appear across artifact types: 49 of the 90 datasets where leakage applies never address train/test leakage, and more than a quarter make no statement about availability. Among the explanatory and observational artifacts, reasoning data has appeared only since 2024 and is dominated by model-generated annotation, while execution traces rest on three deep-coded contributions; nor is the data well understood, as no quality attribute has been the primary
subject of more than fourteen studies. Availability shows a different pattern: executable artifacts, although the most expensive to build and difficult to reconstruct, are tied with patch datasets for the lowest public-availability rate among the major types, at 12 of 24. Reasoning and trace data are also among the least represented artifact types and have received the least direct study, despite their growing importance for AI-driven security.

\paragraph{Contributions.}
This study makes five contributions:
\begin{itemize}
\item A \textbf{data-centric taxonomy} of vulnerability data (Section~\ref{sec:tax}) that cleanly
  separates what the data is, how good it is, and what it is for, with item-level definitions designed for consistent, auditable coding.
\item A \textbf{systematically assembled, auditable corpus} and a deep-coding of 111 anchor papers in which every affirmative rubric-graded value is backed by a verbatim span, absence values having been checked by targeted full-text scans, together with the harvesting, coding, and analysis
  pipeline that produced them---reproducible end to end except for the historical LLM-assisted
  screening decisions, which are inspectable per record but not re-executable
  (Appendix~\ref{app:protocol})
  (Section~\ref{survey-methodology}).
\item A \textbf{quality synthesis} (Section~\ref{survey-result-data-quality-rq2-the-core}) that, for
  each attribute, separates how much it has been \emph{studied} from how well datasets actually
  \emph{achieve} it---revealing patterns that a single per-dataset score would obscure---and shows that the
  six attributes constrain one another rather than varying freely.
\item An \textbf{artifact$\times$quality matrix} (Figure~\ref{fig:matrix}) that shows where specific quality weaknesses occur across artifact types.
\item A statement of the field's \textbf{structural limitations} (Section~\ref{sec:limitations})---seven things existing data cannot do, including three failure modes the corpus cannot
  currently support work on---and an agenda for evidence-rich vulnerability data built from them
  (Section~\ref{sec:future}).
\end{itemize}

\noindent The rest of the study is organized as follows. Section~\ref{sec:related} reviews the three
eras of vulnerability analysis, introduces the running example, and positions this study against
related reviews. Section~\ref{survey-methodology} describes the literature search, screening, and coding protocol.
Section~\ref{sec:tax} presents the taxonomy.
Section~\ref{paper-attribution-and-survey-results} reports the results: corpus distributions, the artifact types with T2--T4 in demonstrative order followed by reasoning and traces as explanatory and observational categories, the per-attribute quality findings that form the study's core,
consumption trends, and the synthesis matrix. Section~\ref{sec:discussion} distils current good practice attribute by attribute, identifies the limitations that remain, and develops future directions for addressing them. Section~\ref{threats-to-validity} states the threats to validity, and Section~\ref{sec:conclusion} concludes.

\section{Background and Related Work}\label{sec:related}

\subsection{Vulnerability Analysis and Its Data}
Automated vulnerability analysis has passed through three overlapping eras, and each was defined as
much by the data it could afford to collect as by the technique it applied.

\textbf{The program-analysis era} relied on hand-crafted, largely \emph{synthetic} suites such as SARD
and Juliet. Their appeal was that ground-truth labels were available by construction for the intended synthetic weaknesses: a sample carried a weakness because it had been written to. Their cost became apparent only when
the same detectors were run elsewhere. Artificially generated vulnerabilities proved easier to detect
than real ones, and detector rankings inverted between benchmark
suites~\cite{pseedevaluatingcomparingmemory}, while per-category recall ranged from 0\% to 100\%,
making any single headline accuracy figure
meaningless~\cite{pseedpreliminarystudyopen}. The labels were reliable by construction, but the resulting data was less realistic.

\textbf{The deep-learning era} changed this tradeoff by using security-fix commits as labels. Treating code touched by a security-fix commit as a proxy for a vulnerability label turns an expert-judgment problem into a data-engineering one, and this is what enabled supervised vulnerability datasets at much larger scale. Devign~\cite{pseeddevigneffectivevulnerability}, Big-Vul~\cite{pseedbigvul},
CVEfixes~\cite{pseedcvefixesautomatedcollection}, and PatchDB~\cite{pseedpatchdblargescale} are the
canonical results, and the corpora built this way remain the field's training base. However, this labeling strategy also introduced systematic errors. CVEfixes notes that when a commit is classified as vulnerable, every function it touches may be labeled vulnerable even though not every changed function is~\cite{pseedcvefixesautomatedcollection}. The consequences became clear when detectors trained on these corpora were evaluated on more realistic data and lost more than half of their performance~\cite{pseedreveal,pseedprimevul}.

\textbf{The LLM-and-agent era} adds two data types the earlier eras had no need for. The first is
corpora of AI-\emph{generated} code whose security must be judged rather than assumed---a reversal in
which the model is the producer of the data, not only its
consumer~\cite{p1520,p850,p1222}. The second is executable benchmarks that probe what coding agents
\emph{do}: not the text they emit but the actions they take against a real or emulated
system~\cite{p1259,pseedagentdojodynamicenvironment,p1109,p863}. This era also breaks an assumption
the previous two shared. When the training corpus is undisclosed, a test set can no longer be assumed
unseen, so pretraining contamination introduces an additional threat that is often difficult to measure directly because training corpora are undisclosed~\cite{p378}.

Across all three eras, the labeling method a technique could afford shaped what data could be collected, which in turn constrained what the technique could demonstrate. This dependence motivates our study.

\subsection{What Counts as Evidence: A Running Example}\label{sec:example}
The taxonomy in Section~\ref{sec:tax} distinguishes six artifact types, and the distinction is easiest
to see on a single vulnerability. Figure~\ref{fig:example} shows an illustrative out-of-bounds write
of the kind that dominates the C/C++ corpora studied here---a length field read from attacker
input and passed to \texttt{memcpy} without being related to the destination's capacity. We use a
composite rather than a specific CVE so that every claim below follows from the code shown.

\begin{figure}[t]
\centering
%% verbatim fills the text block, so \centering alone will not center it;
%% wrapping it in a sized minipage centers the listing as a unit.
\begin{minipage}{0.78\linewidth}
\small
\begin{verbatim}
int parse_record(const char *src, size_t n) {
    char   buf[64];
    size_t len = read_len(src);  /* attacker-controlled */
-   memcpy(buf, src + 4, len);   /* no bound on len */
+   if (len > sizeof(buf)) return -1;
+   memcpy(buf, src + 4, len);
    return handle(buf, n);
}
\end{verbatim}
\end{minipage}
\caption{An illustrative out-of-bounds write and its fix. The same vulnerability can be represented as
any of the six artifact types in Table~\ref{tab:ladder}, which differ in what they let a consumer
conclude.}
\label{fig:example}
\end{figure}

\begin{table}[t]
\caption{One vulnerability, six artifacts. Rows run in the demonstrative-strength order defined in Section~\ref{sec:aspect-a}, which rises
to executable evidence at T4: each row down to that point settles more of what a consumer would
otherwise have to take on trust. Traces are placed last as the record of an execution rather than as
a stronger demonstration; they presuppose a triggering input and require interpretation. The ordering
is not monotonic in explanatory content, nor in production cost.}
\label{tab:ladder}
\small
\begin{tabularx}{\linewidth}{@{}p{1.5cm}p{3.1cm}XX@{}}
\toprule
Artifact & For our example & Establishes & Leaves open \\
\midrule
T2 metadata & CWE-787; severity; file and line & that a category was assigned & nothing about the
code; CWE classes overlap semantically \\
T1 code & \texttt{parse\_record} labeled vulnerable & that someone asserted this function is
vulnerable & which statement, why, whether it is reachable, whether the label is right \\
T5 reasoning & ``\texttt{len} is attacker-controlled and unrelated to \texttt{sizeof(buf)}'' & why
someone judged the code wrong & whether the judgment is correct; who or what made it \\
T3 patch & the two-line diff & that a developer with commit rights changed this code to address it &
whether every changed line is security-relevant \\
T4 test/PoC & an input with \texttt{len} $=$ 65{,}535 & that the triggering behaviour is reachable in the tested environment, \emph{by demonstration} & whether other paths also reach it \\
T6 trace & sanitizer log: write of 65{,}535 bytes at offset 64 & what actually happened at run
time & requires that a triggering input already exist; needs interpreting \\
\bottomrule
\end{tabularx}
\end{table}

Read down Table~\ref{tab:ladder}, the artifacts move from \emph{asserting} that a vulnerability
exists to \emph{demonstrating} that it does. Production cost broadly rises with that movement but not
row by row---a patch is cheaper to obtain than an expert-written rationale, and both are far cheaper
than a working exploit. A patch-derived T1 label can cost little more than extracting a diff. A T4 exploit costs security expertise and a working build
environment---which is why, after processing roughly 2{,}500 sources and 448 CVEs, one benchmark
retained just 23 instances with functioning exploits~\cite{p778}, and why 78.9\% of CVEs have no publicly available proof-of-concept~\cite{p898}.

In practice, the most widely available vulnerability data is therefore concentrated in the cheaper, weaker-evidence artifact types. One row carries a property none of the others does: an executable artifact carries its own check. An exploit that runs is simultaneously the artifact and independently checkable evidence that the behaviour it triggers is reachable, which is why proofs of vulnerability are hard to generate but easy to
verify~\cite{p2109}. The property does not extend downward---a trace records what happened but does
not by itself establish that what happened was a vulnerability, and needs interpreting before it
labels anything. Nothing analogous is available for a mined function, whose label can only ever be
as good as the procedure that assigned it.

This ladder is the organizing idea of the study. It supplies the row order of our artifact taxonomy
(Section~\ref{sec:tax}), the ordering of the artifact discussion
(Section~\ref{survey-result-data-artifacts-rq1}), and the interpretation of the quality matrix
(Section~\ref{survey-result-data-quality-rq2-the-core}), where executable artifacts (T4) turn out to be the only well-populated type for which majorities satisfy both realism and label evidence---and, simultaneously, among the least often shared. The two results have different causes: a run checks the triggered behaviour directly, while the realism result reflects authentic targets with their deployment context retained. We return to the example where it clarifies a result.

\subsection{Related studies}
Most studies of this area organize it by \emph{technique} (detection, repair, generation) or by
\emph{threat}, and treat the data as background. A smaller group takes the data itself as the unit
of analysis, and it is against that group that our contribution has to be stated.

Studies of vulnerability datasets as such exist. Karim and Akter~\cite{bkarim2026vulnerabil} study vulnerability datasets for ML-, DL-, and LLM-based analysis, covering dataset construction, granularity, coverage, label quality, duplication, generalization, and reproducibility. We state our contribution relative to that class of work rather than to any technique study. Two things distinguish it. Our unit
of analysis is the full range of evidence a vulnerability claim can rest on, which
puts metadata, patches, executable proofs, reasoning, and execution traces in scope
alongside code samples, and orders them by what each \emph{establishes} rather than by
what each contains. And our synthesis is coded rather than narrative: a shortlist read against a fixed schema in which every affirmative rubric-graded value carries a verbatim span, which is
what lets us report, per attribute, how much it has been \emph{studied} separately
from how well datasets \emph{achieve} it.

Two further reviews come close to our stance from other directions, and drawing the distinction
sharpens the contribution again. He et
al.~\cite{p636} review quality issues in LLMs for code across 114 primary studies, tracing generation
defects to imperfections in the training corpus through a nine-dimension taxonomy of code-quality
issues. Their object is \emph{general} code quality---correctness and maintainability, with security
one dimension among many---and the \emph{generation} pipeline; ours is \emph{security} vulnerability
data specifically, across its full lifecycle from detection through patching, validation, and agent
behavior, organized by the artifact/quality/use structure rather than a data-to-defect causal chain.
Young and Moody~\cite{p232} invert the usual framing as we do, treating datasets as the unit of
analysis, but for a single sub-domain: thirteen malicious-code \emph{prompt} corpora for refusal
evaluation, which they find were built under thirteen incompatible protocols and for which they call
for standardized reporting---a conclusion our broader study corroborates and generalizes
(Section~\ref{sec:discussion}). Where they treat one artifact type in depth, we span the six artifact
types and six quality attributes of the vulnerability-data landscape.

A third group of reviews is adjacent in subject but not in stance. A systematic review of automated
exploit and security test generation catalogues the techniques that would produce the T4 artifacts of
Table~\ref{tab:ladder}, and reports that few of the studies it studies release usable
tools~\cite{p1201}---an availability finding we corroborate at corpus scale
(Section~\ref{availability}), but one reached while studying techniques rather than data. More
broadly, studies of DL- and LLM-based detection, program repair, and exploit generation catalog
methods and cite datasets in passing. Dataset-centric studies exist, as the comparison above sets
out; what we have not found is one that spans the full artifact range from metadata to execution
traces, or that reports, per attribute, how much it has been \emph{studied} separately from how
well datasets \emph{achieve} it. That is the gap this study fills.

\section{Methodology}\label{survey-methodology}

This study is organized around three research questions about the
\textbf{data} underlying software vulnerability analysis, rather than
the analysis techniques themselves:

\begin{itemize}
\tightlist
\item
  \textbf{RQ1 (Artifact).} Which vulnerability-data artifact types receive attention in the literature, and what artifacts and structural granularities are represented among the deep-coded contributions?
\item
  \textbf{RQ2 (Quality).} How good is that data---along realism, label evidence, scale, diversity, leakage/contamination, and availability---and where are the systematic weaknesses?
\item
  \textbf{RQ3 (Consumption).} Which analysis tasks and technique
  paradigms consume the data, and how has that shifted from traditional
  analysis through deep learning to LLMs and coding agents?
\end{itemize}

We answer these with a documented pipeline: a three-phase literature search, a
data-centric screening filter, a two-tier reading protocol, and a taxonomy-driven
deep-coding scheme in which every affirmative rubric-graded value is backed by a verbatim span from the source, absence values being checked by targeted full-text scans. Harvesting, coding, and analysis re-execute from the released
scripts; one stage, the LLM-assisted screening of 668 ambiguous abstracts, does not, and Appendix~\ref{app:protocol} says why. That stage no longer decides any exclusion on its own: every record it excluded was re-examined in a re-screen of the complete excluded set distributed across three authors (Section~\ref{threats-to-validity}), so the final exclusions rest on the documented human pass. Each stage is described below;
Appendix~\ref{app:protocol} records the queries as issued, the retrieval parameters,
the de-duplication rule, and how the screening dispositions partition the candidate
pool, and the queries, screening prompts, coded records, and analysis scripts are
released with the corpus.

\subsection{Literature Search}\label{literature-search}

\textbf{Phase 1---topic harvesting.} We queried arXiv, DBLP, and---best effort---Semantic
Scholar, enriching results through OpenAlex, across six topical families that jointly span the
artifact taxonomy:
(i) vulnerability datasets and benchmarks; (ii) vulnerability labeling, CWE
assignment, and label quality; (iii) security patches and vulnerability-fix
mining; (iv) proof-of-concept, exploit-generation, and fuzzing benchmarks;
(v) vulnerability reasoning, explanation, and root-cause data; and (vi) the
security of AI-generated code and of coding agents. The harvester carries these six areas as four
operational tags---data, AI-for-security, AI-generated-code security, and agent behaviour---since
families (i)--(iii) share a query vocabulary; Appendix~\ref{app:protocol} lists every query under
its tag. Within each family we combined
a domain term (\emph{vulnerability}, \emph{CWE}, \emph{CVE}, \emph{exploit},
\emph{secure code}) with a data term (\emph{dataset}, \emph{benchmark},
\emph{corpus}, \emph{labeled}, \emph{ground truth}), using stemming so that, for
example, \texttt{vulnerab*} retains all inflections. The primary window is
\textbf{2016--2026}, the period over which learning-based and then LLM-based
vulnerability analysis grew, with foundational datasets from any earlier year
admitted explicitly in Phase 2.

\textbf{Phase 2---named-artifact seeding.} Because foundational datasets are
cited far more often than they are retrieved by topical search, we seeded the
corpus directly with canonical resources---SARD, Juliet, Devign, Big-Vul,
CVEfixes, PatchDB, ReVeal, DiverseVul, VulDeePecker, LineVul, Magma, and others---together with their citing and cited neighborhoods. This corrected a marked
early bias toward post-2021 work.

\textbf{Phase 3---enrichment and backfill.} Candidate records were enriched with
venue, year, citation count, and abstract via OpenAlex, Unpaywall, and Semantic
Scholar, then extended by backward citation chasing from the seeds. Three
additional papers were added on expert recommendation late in the process and
coded identically.

The three phases yielded \textbf{2{,}112 candidate records}. The sources contribute
1{,}194 from arXiv, 776 from DBLP, 236 from a supplemental, recall-oriented search carried out independently by two of the authors with LLM assistance, recorded as such in the released candidate table, 31 curated seeds, and 3 expert recommendations; these sum to 2{,}240, from
which de-duplication by DOI and normalized title removed 128.

\subsection{Screening: a data-centric filter}\label{screening-a-data-centric-filter}

A candidate is \textbf{included} only if it satisfies the inclusion criterion (\textbf{IC}) and
no exclusion criterion (\textbf{EC}).

\begin{itemize}
\tightlist
\item \textbf{IC1.} The paper introduces or produces new vulnerability data,
  builds an approach that generates such data, or reports a finding about the
  quality of existing vulnerability data.
\item \textbf{EC1.} The paper merely \emph{consumes} an existing dataset to
  evaluate a model or technique and contributes nothing to the data itself---a
  model comparison on Big-Vul that neither extends nor characterizes it, for
  instance. This is the exclusion that most sharply distinguishes our corpus from
  a technique study's.
\item \textbf{EC2.} The paper is a pure attack or defense technique with no
  released data artifact. Such work is cited as context where it motivates a data
  need, but is not classified in the matrix.
\item \textbf{EC3.} The subject is outside software security---other modalities
  (vision, generic NLP), generic model jailbreaking with no software context, or
  content abuse.
\item \textbf{EC4.} The record is not a research contribution (editorial, index
  entry, duplicate preprint of an included paper).
\end{itemize}

Screening combined automated rules over title, abstract, and venue metadata
(1{,}171 decisions) with LLM-assisted abstract screening for cases the rules left
ambiguous (668), independent screening by two of the authors (236), and manual
curation of seeds (34). Each candidate received exactly one of these dispositions, and the four counts sum to 2{,}109; the remaining three are the expert recommendations added after screening had run and coded directly.
Table~\ref{tab:flow} gives the resulting flow. Every
excluded record is retained with its reason, so the filter is auditable rather
than merely described; six records remain unresolved and are treated as excluded.

\begin{table}[t]
\caption{Screening flow. Excluded records are retained with their reason.}
\label{tab:flow}
\small
\begin{tabular}{lr}
\toprule
Stage & Records \\
\midrule
Retrieved and de-duplicated (Phases 1--3) & 2{,}112 \\
\quad excluded: no software-vulnerability or AI-coding-security signal (EC3) & 339 \\
\quad excluded: other modality or domain (EC3) & 128 \\
\quad excluded: offensive-security, CTF, or pentest focus (EC2) & 14 \\
\quad excluded: generic jailbreaking, adversarial ML, content abuse (EC3) & 45 \\
\quad excluded: other (EC1, EC4, refinement passes) & 58 \\
\quad unresolved, treated as excluded & 6 \\
\midrule
\textbf{Included} & \textbf{1{,}522} \\
\quad of which deep-coded (Tier 1) & 111 \\
\bottomrule
\end{tabular}
\end{table}

The included set is heavily weighted toward recent and unrefereed work:
\textbf{813 of the 1{,}522 (53\%) are preprints}, 219 (14\%) appear in other venues, 57 (4\%) in indexed journals or top-tier conferences, and 433 (28\%)
carry no venue label we could resolve. We cannot cleanly separate how much of
that skew is a property of the area, which is publishing faster than peer review absorbs it, from how much is a property of a search that reaches arXiv more
readily than paywalled venues; the large unlabeled fraction limits what the
composition can be taken to show either way, and we draw no conclusion from it.
It does bear on how much weight any single included record carries, and it is one
reason the Tier-1 shortlist is selected rather than sampled: the deep-coded set is
better refereed than the corpus it is drawn from, with 52 of its 111 papers in refereed venues against 35 preprints.

\subsection{Two-tier reading protocol}\label{two-tier-reading-protocol}

We read the corpus in two tiers, and the tiers answer different questions. The
per-attribute quality findings (RQ2) require a paper's dataset-construction
section read in full, so they rest on the deep-coded Tier 1. The distributional picture of where the literature's attention sits (RQ1) is carried by title, abstract, and metadata across the whole corpus, which Tier 2 supplies at a scale full reading could not reach. Consumption (RQ3) cannot be: task and technique are judgements about how a dataset is used, which we could only make by reading, so Section~\ref{survey-result-consumption-and-evaluation-rq3} characterises the anchor set and says so rather than projecting it onto the literature. Each tier is therefore matched to the evidence its question needs, and no aggregate mixes the two.

\textbf{Tier 1 (deep coding), 111 papers}, read in full and coded on every taxonomy
attribute. These establish Section~\ref{paper-attribution-and-survey-results}'s
findings and the matrix (Figure~\ref{fig:matrix}). A paper enters Tier 1 if it
meets any of four criteria: (T1a) it is a foundational dataset that later work
builds on, identified by seeding rather than by citation threshold; (T1b) it is
among the most-cited recent contributions within its artifact type, applied per
type so that no type is crowded out by the volume of code-sample work; (T1c) its
contribution is a \emph{finding about} data quality, admitted regardless of venue
or citation count, since such papers are the study's primary evidence; or (T1d)
it is the only, or one of very few, representatives of an artifact type or domain
that would otherwise be unrepresented. Criterion T1c is why the shortlist includes
several lightly cited preprints, and T1d is why reasoning and trace data appear at
all.

\textbf{Tier 2 (light coding)}, the remaining papers, tagged for artifact type,
task, and technique to produce the distribution figures and cluster citations,
with quality attributes recorded where the abstract supports them.

\subsection{Deep-coding scheme}\label{deep-coding-scheme}

Each Tier-1 paper is coded against a fixed extraction form derived from the
taxonomy (Section~\ref{sec:tax}): identity fields (title, venue, year); one
primary artifact type (T1--T6) with secondary types recorded separately;
granularity and provenance; the six quality attributes; the consuming task and
technique; two scope flags (dataset/study/both, core/context); a free-text
\emph{data gap}; and a set of verbatim evidence spans. Values are drawn from
closed enumerations wherever a judgment is graded, so that records are comparable
and machine-checkable; the schema is released with the corpus.

Two conventions established during coding are worth stating. A \texttt{study}
paper is scored on the \emph{data it examines}---fully when it studies one
dataset, but with realism, label evidence, and leakage left
\emph{not-assessed}, and scale and diversity carrying the review's own scope, when
it spans many. And an analysis lens, \texttt{studies\_attributes}, records which
quality attributes a paper produces a \emph{finding} about, so that
Section~\ref{survey-result-data-quality-rq2-the-core} can report how much each
attribute has been \textbf{studied} separately from how well datasets
\textbf{achieve} it. Both conventions arose from disagreements during calibration
and are written into the codebook as decision rules with worked examples.

\textbf{Calibration.} Before coding at scale, all coders independently coded a
common set of papers and compared results. Two systematic ambiguities surfaced---the treatment of \texttt{study} papers and the scope of \texttt{context}---and
were resolved by adding the rules above. Enumerated values were extended twice
during coding, when a value observed in the corpus had no faithful code:
\texttt{tool\_derived} for labels assigned by a static analyzer with no human or
execution check, and \texttt{contamination\_controlled} for datasets that guard
against pretraining exposure by construction rather than by a split.

\textbf{AI-assisted coding with mandatory evidence.} Each record was first drafted by an LLM from
the paper's full text, and all 111 were then verified by the authors. For 80 records an author checked every field against the source; the remaining 31, merged later from a second extraction
pipeline, were screened for conflicts against the codebook, their own evidence spans, and an
independent earlier coding, and every flag raised was adjudicated by three authors
(Section~\ref{threats-to-validity}). The governing rule is anti-hallucination: no affirmative rubric-graded value is recorded unless a \textbf{verbatim span} from the paper supports it, and absence values are checked by targeted full-text scans. All 111 records carry evidence spans, 729 in total, and numeric values were
additionally checked mechanically against the source text. What that verification
found, and what it implies for the reliability of the results, is reported in
Section~\ref{threats-to-validity}.

\section{Taxonomy of Vulnerability Data}\label{sec:tax}

Our taxonomy is deliberately \emph{dataset-centric}: it describes three aspects of a
vulnerability-data artifact---what it \emph{is} (Form, A), how \emph{good} it is (Quality, B), and
what it is \emph{for} (Use, C)---with no attribute appearing in more than one aspect
(Table~\ref{tab:taxonomy}).

Three principles govern it. \textbf{Separation:} an attribute belongs to exactly one aspect, so that
a claim about quality cannot smuggle in a claim about form. \textbf{Codability:} every graded attribute has a closed vocabulary whose items are defined operationally, to reduce discretion and make disagreements auditable across readers; where a property resists that, we record a reported figure
instead of a grade. \textbf{Consumer-relevance:} an attribute earns its place only if a downstream
consumer's trust depends on it. Metrics, model architectures, and evaluation protocols are properties
of \emph{studies}, not of data, and are excluded on that ground.

Those principles resolved three concrete design choices. The \emph{production method} of a
vulnerability---mined from history, injected by a tool, or synthesised outright---is folded into
\emph{realism} rather than given a separate ``how-made'' axis, because how a vulnerability was
produced matters only insofar as it changes how real it is. \emph{Provenance} is a different
question and keeps its own attribute in Aspect~A: it records who or what authored the surrounding
code, human or model, which cuts across realism rather than determining it. The evidence ladder and reproducibility are
treated as cross-cutting analytical \emph{lenses} rather than axes, since both are readings of the
type and quality columns rather than independent properties. And evaluation \emph{metric} is dropped
entirely.

The taxonomy is not the first for data quality in this space, and the differences are deliberate. A
recent review of quality issues in code LLMs organizes generated-code defects along nine
dimensions~\cite{p636}; that taxonomy is oriented to \emph{general} code quality, with security one
dimension among several, and to the generation pipeline. Ours is oriented to security data
specifically and to its whole lifecycle, from mining and labeling through validation and agent
behavior. General data-quality frameworks contribute the vocabulary of completeness, consistency, and
accuracy, but they are domain-neutral and do not distinguish an artifact that asserts a vulnerability
from one that demonstrates it, the distinction this study is organized around.

\subsection{Derivation}
We derived the taxonomy in two passes over the corpus. \textbf{Identification.} Reading the papers, we
recorded candidate attributes (``real vs.\ synthetic,'' ``CWE label source,'' ``PoC available,''
\ldots) and their values, keeping the three questions---form, quality, use---separate from the start.
\textbf{Generalization.} We then generalized to a reusable set and resolved overlaps.

Three candidate attributes were rejected in the second pass, and the reasons illustrate the
principles. \emph{Evaluation metric} was dropped as a property of a study rather than of data.
\emph{How-produced} was folded into realism because every value it could take mapped onto a realism
value, making it redundant. \emph{Reproducibility} was demoted to a lens because it is a function of
availability, leakage handling, and label provenance jointly rather than an independent property---as
the reproducibility funnel in Section~\ref{threats-to-validity} shows, it is a downstream consequence
of the other three.

Two values were added to the vocabulary \emph{during} coding, when the corpus contained cases no
existing item described faithfully. \emph{Tool-derived} labels, assigned by a static analyzer with no human or execution check, are none of verified, patch-derived, or description-derived, yet they
recur across the corpus and are criticized from within it: Devign rejects prior datasets because the
labels ``are generated by statistic analyzers which are not
accurate''~\cite{pseeddevigneffectivevulnerability}, and both the label-error and Mono studies trace
noise to the same source~\cite{p1420,p951}. \emph{Contamination-controlled} covers datasets with no
train/test split to speak of that nonetheless guard against pretraining exposure by construction, a
category that did not exist before LLM consumers did. Both additions are disclosed here because they
postdate the schema a reader would otherwise assume was fixed in advance.

\begin{table}[t]
\caption{The data-centric taxonomy: three aspects, their attributes, and item vocabularies. Quality
items are defined in Section~\ref{sec:quality-items}.}
\label{tab:taxonomy}
\footnotesize
\begin{tabularx}{\linewidth}{@{}p{1.7cm}p{2.05cm}XX@{}}
\toprule
Aspect & Attribute & Items & Description \\
\midrule
\textbf{A. Data Artifact} \textit{(form)} & Artifact type & T1 Code samples & vulnerable/benign code units---the object under study \\
 & & T2 Metadata & CVE/CWE labels, vulnerability type, severity, location, description \\
 & & T3 Patches \& fixes & remediating code changes (fix diffs/commits) \\
 & & T4 Tests, PoCs \& exploits & executable inputs that trigger/demonstrate the vulnerability \\
 & & T5 Reasoning & root-cause explanations, rationale, chain-of-thought, exemplars \\
 & & T6 Traces \& logs & execution traces, crash/error logs, vulnerability or agent trajectories \\
 & Granularity & statement/slice $\cdot$ function $\cdot$ file $\cdot$ module $\cdot$ repository $\cdot$ n/a & structural scope of the artifact \\
 & Provenance & human $\cdot$ AI-generated $\cdot$ mixed $\cdot$ n/a & who or what authored the code \\
\midrule
\textbf{B. Data Quality} \textit{(how good)} & Realism & synthetic $\cdot$ injected/generated $\cdot$ mined/semi-real $\cdot$ real-world $\cdot$ not assessed & authenticity vs.\ real-world vulnerabilities \\
 & Label evidence & verified $\cdot$ patch-derived $\cdot$ tool-derived $\cdot$ description-derived $\cdot$ unlabeled $\cdot$ not assessed & how labels/locations were obtained, hence their trustworthiness \\
 & Scale & number of samples (with unit) & quantity---a figure, not a judgment \\
 & Diversity & CWE $\cdot$ project $\cdot$ language coverage $\cdot$ class balance & breadth and vulnerable:benign balance \\
 & Leakage \& contamination & random $\cdot$ dedup $\cdot$ project-disjoint $\cdot$ chronological $\cdot$ contamination-controlled $\cdot$ not addressed $\cdot$ n/a & train/test independence, duplication, pretraining exposure \\
 & Availability & public $\cdot$ on-request $\cdot$ unavailable $\cdot$ not reported & whether the dataset can be obtained and reused \\
\midrule
\textbf{C. Consumption} \textit{(use)} & Consuming task & detection $\cdot$ localization $\cdot$ classification $\cdot$ explanation $\cdot$ repair $\cdot$ validation $\cdot$ behavior/safety vetting & the analysis task the data serves \\
 & Consuming technique & traditional $\cdot$ deep learning $\cdot$ fine-tuning $\cdot$ LLM prompting $\cdot$ RAG $\cdot$ agentic $\cdot$ benchmark & technique paradigm consuming the data \\
\bottomrule
\end{tabularx}
\end{table}

\subsection{Aspect A: form, and why the types are ordered}\label{sec:aspect-a}
Each artifact is assigned \emph{one} primary type, so that the status-quo matrix
(Figure~\ref{fig:matrix}) has clean rows; bundled artifacts are recorded separately but do not
change the primary type.

The six types are not an unordered list. As Section~\ref{sec:example} shows on a single vulnerability,
they differ in what they let a consumer conclude, and that difference is systematic: metadata asserts a category, code samples assert that a unit is vulnerable, reasoning asserts
\emph{why} it is, patches localize where a developer thought the flaw lived, and executable
artifacts demonstrate that the behaviour is reachable. Traces sit apart: they record what happened
when something ran, which is observational rather than demonstrative, and they presuppose that a
trigger already exists. We place them last in the row order for that reason, not because a trajectory
is stronger evidence than a working exploit. We order the types by that
criterion, evidence strength, and use the same order for the rows of the quality matrix and the
sequence of Section~\ref{survey-result-data-artifacts-rq1}.

\textbf{Placing reasoning.} Reasoning appears between code samples and patches for presentation, but it is not a stronger rung on the demonstrative axis. The strict demonstrative ordering runs from metadata and code-sample assertions through patch evidence to executable demonstration; reasoning and traces supply explanatory and observational dimensions alongside it rather than positions within it. A rationale adds explanatory content a bare label lacks, but it remains an \emph{assertion}, and nothing in it makes the flaw more demonstrably real---in our corpus most
reasoning is model-generated, so the account and the label often share an author
(Section~\ref{res:t5}). A patch outranks it because a patch is the residue of an action taken by
someone with commit rights, which localizes the flaw independently of any annotator's account.

The ladder measures demonstrative strength only. Ranked instead by
\emph{explanatory} content, reasoning would sit at the top and a trace near the
bottom, since a raw execution record explains nothing until it is interpreted; the T5
row should be read against whichever axis a given question calls for. The ordering is a claim about what each artifact
\emph{establishes}, not about how useful it is; a large mined corpus is more useful than a single
exploit for training a detector, and weaker as evidence about any one sample.

Granularity records the structural scope of the artifact and matters more than it first appears,
because it is the axis along which realism is traded for labelability
(Section~\ref{realism}). Provenance records whether the code was written by people, generated by a
model, or both---a distinction that barely existed when the earliest datasets here were built and is
now central, since a consumer cannot otherwise tell mined code from generated code.

\subsection{Aspect B: quality-item definitions}\label{sec:quality-items}
Four quality attributes are graded against closed vocabularies designed to support consistent coding across readers. Scale and diversity are recorded as reported figures rather than
judgments, because sensible thresholds differ by artifact type.
\begin{itemize}
\item \textbf{Realism.} \emph{Synthetic}: hand-crafted or templated examples not from real software
  (SARD, Juliet). \emph{Injected/generated}: real code into which a vulnerability was inserted by a
  rule, tool, or model (LAVA, VulGen~\cite{pseedvulgenrealisticvulnerable}). \emph{Mined/semi-real}:
  collected from real projects or CVEs at function or file granularity, but simplified or carrying
  imperfect labels. \emph{Real-world}: authentic, full-context vulnerabilities from real projects.
  \emph{Not assessed}: the paper spans too many datasets for one value to be meaningful.
\item \textbf{Label evidence.} \emph{Verified} (manual review, execution, or a PoC);
  \emph{patch-derived} (inferred from the fix commit---the dominant practice, imperfect for tangled
  commits); \emph{tool-derived} (a static analyzer's output, no human or execution check);
  \emph{description-derived} (from CVE/NVD text or keywords, no code check); \emph{unlabeled};
  \emph{not assessed}. The ordering of the first four is a cost ordering as much as a quality one
  (Section~\ref{label-correctness}). The attribute records the \emph{evidential basis} of a label, how it was obtained and what independently checked it, rather than measuring whether the label is
  right, which no coding of a paper could establish. \emph{Verified} accordingly means an independent
  check was applied, by execution or by human review; it does not certify the assigned CWE, the root
  cause, or the absence of other flaws in the sample.
\item \textbf{Scale.} The number of samples with its unit (e.g., 188k functions; 1{,}740 CVEs; 318
  tasks). Deliberately not bucketed into small/large, since sensible boundaries differ by artifact
  type; comparison is made within a type.
\item \textbf{Diversity.} Coverage as reported, on four facets: number/range of CWEs, distinct
  projects, programming language(s), and vulnerable:benign class balance.
\item \textbf{Leakage \& contamination.} \emph{Random} (no dedup, so duplicates may span train and
  test); \emph{dedup}; \emph{project-disjoint}; \emph{chronological} (train older, test newer);
  \emph{contamination-controlled} (no split applies, but pretraining exposure is guarded against by
  construction---post-cutoff, proprietary, or mutated data); \emph{not addressed}; or \emph{n/a},
  where no learned model consumes the data and neither concern arises. The last two are distinct and
  the distinction matters: \emph{n/a} says the question does not apply, \emph{not addressed} says it
  applies and went unanswered.
\item \textbf{Availability.} \emph{Public}, \emph{on-request}, \emph{unavailable} (including
  ``released upon publication''), or \emph{not reported}. ``Not reported'' is not the most common value, since most of the datasets we coded are public, but it is five times more common than an explicit refusal.
\end{itemize}

\subsection{Aspect C: consumption}
Consuming task and technique record who the data is for. Both are multi-valued, since a dataset
commonly serves several tasks, and both are recorded from what the paper itself evaluates rather than
from what its data could in principle support. The technique vocabulary is deliberately paradigmatic
rather than architectural---\emph{deep learning} rather than a list of model families---because the
study's question is which \emph{kind} of consumer a dataset was built for, and that is what has
shifted across the three eras of Section~\ref{sec:related}.

\subsection{Coding rules}
\textbf{One primary type.} Big-Vul bundles code, metadata, and patches but contributes a labeled code
corpus (T1); CVEfixes contributes a fix collection (T3); a CWE-mislabeling study is T2; a
PoC-generation paper is T4. For benchmarks that bundle code with executable verification the
tie-break is the novel contribution: an executable harness makes it T4 (SEC-bench,
CyberGym~\cite{p1109}, VulnRepairEval~\cite{p778}); a labeled corpus makes it T1
(JavaVulBench~\cite{p412}, Big-Vul).

\textbf{Datasets and studies.} A \emph{dataset} paper introduces a new artifact and is scored on the
quality attributes. A \emph{study} paper analyzes existing data and contributes to one or more quality
attributes; it is coded by the type it studies, and, critically, its quality columns describe the
\emph{data it is about}. A study of one dataset (ReVeal on Chromium and Debian) is scored fully; a
study spanning many datasets records \emph{not-assessed} for realism, label evidence, and leakage
and uses scale and diversity for the review's own scope, since no single value is meaningful across
dozens of datasets. To make ``how much has attribute $X$ been studied'' answerable separately from
``how good is $X$,'' we additionally record, as an analytical lens, which quality attributes each
paper produces a \emph{finding} about.

\textbf{Scope filter.} A paper enters the taxonomy only if it introduces, generates, or reports a
finding about vulnerability data; one that merely consumes existing data is excluded and cited only
where it motivates a data need (Section~\ref{screening-a-data-centric-filter}).

\textbf{Two conventions} were fixed during coding. Commit- and diff-level artifacts are coded at
\emph{slice} granularity, since a patch is a set of changed statements rather than a whole function.
And leakage is single-valued, so a dataset shipping several strategies at once (JavaVulBench ships
five) is coded by its strongest and discussed individually in
Section~\ref{leakage-and-contamination}.

\textbf{Distinctions to keep clean.} T4 is the \emph{triggering input} (PoC, exploit, test); T6 is the
\emph{record of execution} (trace, crash log, trajectory)---input vs.\ record. T2 is \emph{factual}
annotation (CVE/CWE, severity, location); T5 is \emph{analytical} content (root-cause
reasoning)---what the data is vs.\ why. Artifact types are technique-agnostic; the consuming technique
lives in Aspect C.

\subsection{What the taxonomy does not capture}
Three limits are worth stating, since each bounds a claim made later.

\emph{Realism is a composite, and one of its values reaches into two other attributes.}
\emph{Mined/semi-real} is defined by origin together with either granularity or label quality, both
of which are coded separately. We kept it because these three properties often co-occur in practice and a single
ordered value is what makes the matrix readable, but it does breach the separation principle above,
and any realism figure should be read alongside the decomposition in Section~\ref{realism} rather
than on its own.

\emph{The aspects are separable for coding, not independent in fact.} Assigning realism and label evidence to different columns lets them be coded without contaminating each other, but they are
empirically coupled through the label oracle, as Section~\ref{attributes-not-independent} argues. The
taxonomy is a measurement instrument, not a causal model.

\emph{One primary type per artifact loses information.} A dataset that ships code, patches, and
executable tests is counted once. Secondary types are recorded, but the matrix uses the primary, so
bundled contributions are under-represented in every row but one.

\emph{Two vocabulary items are defined but never instantiated} in our deep-coded set: provenance
\emph{n/a} and label \emph{unlabeled}. We keep them because each is a coherent possibility the schema
should admit. A third, the \emph{explanation} consuming task, was empty until a corpus-wide screen
for reasoning data added five datasets that use it (Section~\ref{res:t5})---a reminder that an
empty cell can record the limits of a search rather than the state of the field.

\section{Paper Attribution and Study Results}\label{paper-attribution-and-survey-results}

\subsection{Distribution by year and artifact type}\label{distribution-by-year-and-artifact-type}

\begin{figure}[t]
\centering
\includegraphics[width=0.6\linewidth]{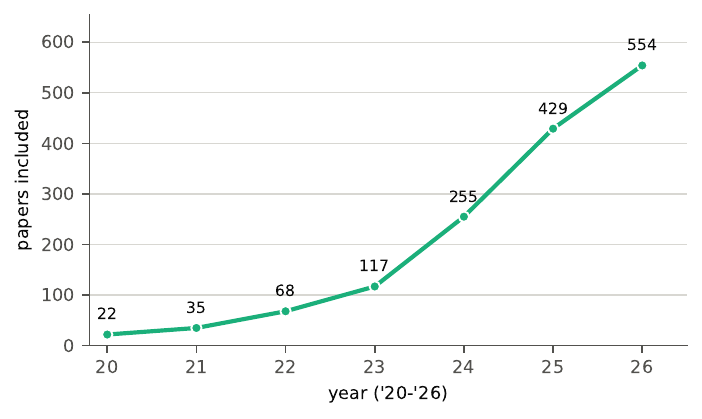}
\caption{Corpus growth, 2020--2026: papers included per year. We plot the total only. The corpus-wide artifact tags come from a keyword tagger whose recall is not constant across the window---it recovers none of the true T4 papers published through 2023 and about two thirds afterwards (Section~\ref{threats-to-validity})---so any series split by type, absolute or proportional, moves partly with detection rather than with output. A yearly total does not: a paper counts once whatever type it is assigned. The per-type tagger counts appear in Table~\ref{tab:t61_type} as descriptive output, and the compositional question is taken up on the hand-coded records in Section~\ref{sec:trajectory}. The 2026 figure covers a partial year and is a lower bound.}
\label{fig:growth}
\end{figure}

The 1{,}522 included papers are heavily concentrated in the last three
years---117 in 2023, 255 in 2024, 429 in 2025, and 554 in 2026, against 22, 35, and 68 in
2020--2022 and a long, thin tail of 42 papers from 2019 or earlier that carry most of the
foundational datasets. The included literature is heavily concentrated in recent years (Figure~\ref{fig:growth}), while the quality problems Section~\ref{survey-result-data-quality-rq2-the-core} documents persist.
The composition of that output appears to shift toward executable artifacts as well, but the
automatic tags cannot establish it: validated against the deep-coded set, the tagger recovers none of
the true T4 papers published through 2023 and two thirds of those from 2024 onward
(Section~\ref{threats-to-validity}), so a measured rise from a 2023 baseline is not separable from a
tagger that misses the artifact entirely in that year. We therefore make no compositional claim from the
corpus-wide tags, and take up the question in Section~\ref{sec:trajectory} on the deep-coded records,
where types are assigned by hand.

By primary artifact type (Table~\ref{tab:t61_type}, which sets the whole corpus beside the
deep-coded set), \textbf{code samples (T1) draw the most literature at 753 papers}, followed by
tests/PoCs/exploits (T4, 376), patches (T3, 156), metadata (T2, 119), traces (T6, 94), and reasoning
(T5, 24), covering all 1{,}522 included papers. These are counts of \emph{papers associated with} an
artifact type, not of artifacts produced: a study of code datasets counts as T1 while contributing no dataset of its own, and a paper
shipping code, patches, and tests together is assigned one primary type. In the deep-coded set, where
we can tell the two apart, 93 of 111 papers contribute a dataset and 18 study data they did not
build; we have no equivalent split for the corpus at large. The distribution therefore measures where
the literature's attention sits, and we do not read it as the composition of what the field
produces. Representative examples of each of the four
largest groups are T1~\cite{p1482,bzhou2024large,p1470,bzhou2025large,p1356,bzhang2022reentrancy,p1449,p1458,brahman2024causal,bcai2023software,byuan2023enhancing,byang2023does,bchakraborty2024revisiting,bsejfia2024improved}, T4~\cite{bali2025crossguard,bli2025guided,bnakanishi2025fuzzing,bcheng2025discoverin,bhajipour2023systematic,bwang2024your,bchen2025secureagen,bsultana2026llms,byang2026safuzz,bchen2026fuzzysql},
T3~\cite{bpearce2023examining,bdefiterodoming2024enhanced,p1410,bzhang2024evaluating,p988,bluo2025exploring,bzhang2024dual,bhan2025dependable}, and T2~\cite{p1295,balshaya2023vulnerabil,bmalhotra2025hybrid,bnana2025weakweb,bokeke2026binary,bxiao2026clvul}. These are auto-assigned tags and should be read as indicative. Within the hand-coded validation set the tagger systematically \emph{under}-counts T4, often assigning executable benchmarks to T1; the magnitude of any corresponding whole-corpus correction is unknown, since that set is purposive (Section~\ref{threats-to-validity}).

\subsection{Categorization by taxonomy}\label{categorization-by-taxonomy}

Each Tier-1 paper is placed at one primary artifact type and
granularity, scored on the six quality attributes, and tagged with its
consuming task and technique (Section~\ref{sec:tax}). Two papers that only consume
existing data were reclassified as \emph{context} and excluded from the
matrix; one duplicate was dropped.

Tables~\ref{tab:t5x_attribution_a} and~\ref{tab:t5x_attribution_b} give the resulting
attribution in full: every deep-coded paper with its granularity, its four graded quality
values, and whether it contributes data, studies it, or both. They are the audit trail for
the aggregates that follow---any figure in this section can be traced back to the rows
that produce it, and any row to the quoted span behind each of its values.

\subsection{Study result: data artifacts (RQ1)}\label{survey-result-data-artifacts-rq1}

Ordering the types by what each establishes---T2 to T4 in demonstrative order, with reasoning and traces following as explanatory and observational categories---from artifacts that
merely \emph{assert} a vulnerability to artifacts that
\emph{demonstrate} one, as Table~\ref{tab:ladder} lays out on the running example of
Section~\ref{sec:example}---organizes the rest of the study.

Among the 111 deep-coded papers (Table~\ref{tab:t61_type}), \textbf{code samples (T1, 55)}~\cite{p1222,p1230,p1264,p1276,p1433,p1494,p1520,p274,p287,p313,p378,p412,p429,p826,p850,p857,p888,p976,pseedassessingimprovingprompting,pseedbigvul,pseeddevigneffectivevulnerability,pseedevaluatingcomparingmemory,pseedexploringimprovingreal,pseedgeneratingrealisticvulnerabiliti,pseedpreliminarystudyopen,pseedprimevul,pseedreveal,pseedsven,pseedvgxlargescale,pseedvulgenrealisticvulnerable,pseedvulscriberexploringrag}
and \textbf{tests/PoCs/exploits (T4, 25)}~\cite{p1001,p1109,p113,p1201,p1259,p1277,p163,p170,p203,p2109,p2110,p2111,p284,p505,p778,p863,p898,pseedagentdojodynamicenvironment,pseedmagmagroundtruth,pseedtoolemu} dominate, with \textbf{patches (T3, 15)}~\cite{p1403,p195,p220,p38,p669,p768,p951,p974,pseedcvefixesautomatedcollection,pseedpatchdblargescale,pseedvulrepairt5based} third; \textbf{metadata (T2, 5)}~\cite{p1420,p232,p636,p801,p899}, \textbf{reasoning (T5, 8)}~\cite{p1035,p803,p1567,p907,p1246,p800,p791,p486}, and \textbf{traces (T6, 3)}~\cite{p766,bbaumann2026chat,p14} are the smallest groups. The T5 and T6 counts are the
most consequential, and they now diverge. Reasoning data has arrived quickly: eight of its datasets are deep-coded here, and every one of them appeared in 2024 or later. Execution and agent traces have begun to follow, on three deep-coded corpora. Corpus-wide the two have diverged sharply in volume---94 papers carry a trace artifact against 24 carrying reasoning data, and 82 of the 94
appeared in 2026---but that counts papers, not released datasets. The artifact that would let a consumer inspect an account of \emph{why} code is considered vulnerable and the one that would show \emph{what an agent did} are both now being produced; what remains scarce is either one released and documented well enough to reuse.

Granularity is dominated by the function (50 of 111), with repositories second (15) as agentic and repository-level benchmarks arrive; patches
and diffs are coded at slice granularity.

\subsubsection{Metadata (T2)}\label{res:t2}
Metadata sits at the bottom of the evidence ladder because it \emph{asserts} a property without
showing anything: a CVE record, a CWE class, a severity score, a location. It is also the most
widely consumed artifact in the field, because every labeled code corpus is metadata attached to
code. Yet only five deep-coded papers treat it as the object of study rather than as a free
input~\cite{p1420,p232,p636,p801,p899}, and of the three that contribute an artifact, none reaches
real-world realism or verified labels (Figure~\ref{fig:matrix}). Yet metadata, despite being inherited by many downstream datasets, has received relatively little direct study.

The reason is that CWE was designed for human communication rather than for machine labels, and its overlapping, language-dependent categories complicate its use as consistent ground truth. Categories overlap semantically---cross-site scripting and server-side
template injection both instantiate CWE-74---the taxonomy varies across languages, and real corpora
carry incomplete labels, so multi-label CWE assignment is unreliable as ground
truth~\cite{p801}. A label that cannot be assigned consistently cannot serve as reliable single-label ground truth.

The cost of that unreliability has been measured once, and the number is large. Studying label
errors in three detectors trained on SARD and FFmpeg+Qemu, Nie et al.\ find that noise costs up to
20.7\% F1 in the worst case, and that denoising recovers 10.4\% on average~\cite{p1420}. The implication is not that the detectors are weak but that label errors can materially distort reported performance. The errors trace to two mechanical sources---static-analysis false
positives and tangled commits---both of which are properties of how vulnerability metadata is
collected rather than of any individual dataset.

At field scale the same defect appears as incommensurability. A systematic review of thirteen
malicious-code prompt corpora finds thirteen construction protocols: each built under different
inclusion criteria, licensed differently, validated against different inter-rater standards, with
per-prompt reliability typically unreported~\cite{p232}. Anyone consolidating across them must
reconstruct each corpus's inclusion criteria from scratch. A companion review of 114 primary studies
traces the same fragmentation into code LLMs, where duplicated samples cause overfitting and
imbalanced data limits domain adaptability, but no end-to-end quality-assurance guidance
exists~\cite{p636}.

The wider corpus is consistent with this: metadata work concentrates on assigning CWE
categories rather than on the quality of the assignment~\cite{bzhang2025enhancing,p1021,p851,bzhang2024empirical,bwang2023enhancing,bshahzad2026theory,bwang2026realsec,bhajipour2024hexacoder,bmosievskiy2026fine,bkharma2026enhancing}.

Metadata is therefore the artifact type whose remedy is most clearly a reporting one. Better
mining will not fix a taxonomy that different groups apply differently; what would fix it is a shared
extraction template and a reporting standard, so that a CWE label carries with it how it was assigned
and by whom.

\subsubsection{Code Samples (T1)}\label{res:t1}
Code samples are the field's default artifact---55 of the 111 deep-coded papers, and 41 of the 93 contributed datasets---and also its least realistic: \emph{one} of those 41 carries authentically
real-world, full-context vulnerabilities (Figure~\ref{fig:matrix}), and the exception is
instructive. Smart-LLaMA-DPO's samples are whole Solidity contracts~\cite{p843}: at module
granularity the deployed unit \emph{is} the sample, so no context is lost in cutting it out. Every
other code-sample dataset slices a function or a file from a larger project, and none of those
reaches real-world realism.

That is not an accident of sampling but a consequence of how such datasets must be built. Real
vulnerable code is rare---PrimeVul's mined corpus runs 6{,}968 vulnerable against 228{,}800 benign
functions~\cite{pseedprimevul}---and confirming it is expensive: building Devign's corpus took a team of four professional
security researchers 600 man-hours of two-round labeling and
cross-verification~\cite{pseeddevigneffectivevulnerability}. To obtain volume at tolerable cost, builders mine fix commits and slice the
result to \emph{function} granularity, which is what makes automatic labeling tractable. But the
same slice discards the calling context, build configuration, and inter-procedural flow in which the
vulnerability actually manifests. The granularity choice that buys scale is precisely the one that
forfeits realism, so T1 datasets are pushed toward the middle of the realism scale by construction.

The two poles show the trade in the open. At the accurate end, SVEN hand-curates 1{,}606 programs
whose labels are trustworthy but which span only nine CWEs~\cite{pseedsven}, and Devign labels
manually across four projects~\cite{pseeddevigneffectivevulnerability}. At the scalable end, Big-Vul
harvests 3{,}754 vulnerabilities across 348 projects by taking every function a fix commit touched
as vulnerable~\cite{pseedbigvul}---cheap, and wrong often enough that later audits put label accuracy
in such corpora below 60\%~\cite{p951}.

Synthesis was the field's attempt to escape the trade, and it has been refined four times without
escaping it. VulGen mines injection patterns and applies them with a learned
localizer~\cite{pseedvulgenrealisticvulnerable}, \emph{yet} is confined to single-statement
injections and bounded by the small seed corpus it learns from. VGX widens the pattern vocabulary to
604 edits drawn from historical fixes and human CWE knowledge~\cite{pseedvgxlargescale},
\emph{yet} those patterns still cannot exceed what the seed data contains. VinJ industrializes the
pipeline to 686k samples at 0.4 seconds each~\cite{p1230}, \emph{yet} only 69\% of what it emits is
actually vulnerable, so roughly a third of the corpus is mislabeled by construction. VulScribeR
replaces patterns with retrieval-augmented LLM generation at about US\$1.88 per thousand
samples~\cite{pseedvulscriberexploringrag}, \emph{yet} the downstream benefit plateaus: past roughly
5k generated samples, detectors stop improving. Most recently, LLM prompting reaches 88\% injection
success on real-world code and finally breaks the single-line ceiling~\cite{pseedexploringimprovingreal},
\emph{yet} success varies sharply by CWE, and over 65\% of vulnerabilities in the largest real-world
corpora require the multi-line edits that earlier generators could not express at all.

The wider corpus reproduces this shape rather than contradicting it: code samples are the
largest group in the screened set by a wide margin~\cite{bnguyen2024context,p803,btang2024vulloc,byu2025preliminar,bhe2025cosefa,bhuang2026steer,bkalouptsoglou2026locvul,btan2026deepdesc,bliu2025have,bkarim2026vulnerabil,bnguyen2026mando,bkoterba2026leveraging,bshu2026evaluating,bchen2026cotvd,bmao2025mpda,blekssays2025llmxcpg,blin2025large,bmirsky2023vulchecker,bafanador2020representa,bchen2023diversevul,bliu2020deepbalanc,bpurba2023software,btang2023csgvd,bshestov2025finetuning,bhu2023large,bdeng2023smart}.

The frontier is therefore not more data but \emph{validated} data. Every generator above argues
realism indirectly, by showing that detectors trained on its output improve, rather than
demonstrating that a generated sample is exploitable. Execution validation would raise T1's label-evidence result; context-preserving mining would raise its realism result. The two cells move for different reasons, and neither substitutes for the other.

\subsubsection{Reasoning (T5)}\label{res:t5}
Reasoning data---root-cause explanations, annotated safety properties,
chain-of-thought traces---is what would let a consumer inspect and evaluate an account of \emph{why} code is considered vulnerable, rather than merely that a label says so. Eight of the deep-coded datasets contribute it, and every one of them was published in 2024 or later~\cite{p1035,p803,p1567,p907,p1246,p800,p791,p486}. This row is not empty;
it is \emph{new}, and it arrived after most of the corpora it annotates.

That timing shapes everything else about it. Reasoning data is almost always an annotation layer over
code that already exists: R2Vul pairs functions mined from NVD fixing commits with a valid and a
flawed structured explanation~\cite{p907}, Vul-R2 extends PrimeVul with chain-of-thought
answers~\cite{p1567}, and one 53{,}693-explanation corpus annotates SeVC, DiverseVul and PrimeVul
together~\cite{p803}. The consequence is visible in the matrix: \textbf{7 of the 8 are mined/semi-real or synthetic}, because a reasoning layer inherits the realism of whatever it is
layered on. Only VADER, which curates its own 174 cases from open-source repositories rather than
annotating an existing corpus, reaches real-world realism~\cite{p791}.

\textbf{Scale and authorship.} Across these datasets, larger scale coincides with less direct human authorship. VADER's 174 cases are written by security experts, each preparing a full report~\cite{p791}. SmartCoder-R1's 7{,}998 samples are expert-\emph{validated} rather than expert-written~\cite{p800}. R2Vul's 18{,}000 are produced by a teacher model and shaped by AI feedback~\cite{p907}, while the 53{,}693-explanation corpus is model-generated throughout~\cite{p803}. This pattern reflects the same cost tradeoff described in Section~\ref{label-correctness}: explanations can be generated cheaply at scale or reviewed more carefully at higher cost.

One paper measures the cost of that trade directly. Smart-LLaMA scores its generated explanations
both automatically and with eight domain PhD students, and reports that the human scores run
\emph{consistently below} the LLM-assigned ones~\cite{p1246}---a quantified gap between what a model
will certify as a good explanation and what an expert will.

Against that, the row's quality profile is unexpectedly strong on the attributes that cost nothing.
Every reasoning dataset for which leakage applies addresses it, and seven of eight are publicly released---the best figures of any artifact type in the matrix
(Figure~\ref{fig:matrix}). These are recent datasets built by teams already aware of contamination,
and they show what a row looks like when its builders inherit current norms rather than a decade of
accumulated practice. The weakness is not hygiene but grounding: only one dataset in the row rests on
authentically real-world code.

The gap the row still leaves is therefore narrower and sharper than an empty row would be.
Explanations have begun to appear in numbers, but the ones produced at scale are model-generated annotations whose
correctness has been checked by another model, and the one corpus with expert-authored reasoning
stops at 174 cases. What is missing is not reasoning data but reasoning data whose \emph{ground truth
is not itself a model output}---which is exactly the check that would catch the undecidable cases of
Section~\ref{label-correctness}, and exactly the check that none of the reasoning corpora we screened performs at scale.

\subsubsection{Patches and Fixes (T3)}\label{res:t3}
Patches occupy the middle of the ladder, and the matrix places them there empirically: 5 of 14 reach real-world realism and 7 of 14 verified labels, better than code samples on both counts and worse
than executable artifacts on both (Figure~\ref{fig:matrix}). A patch is a stronger localization signal than an unsupported label and a weaker one than an executable demonstration: it shows where a developer judged a fix necessary, without establishing that every changed line was security-relevant.

The gap between those two is the patch-derived label, and CVEfixes states the defect from inside the
practice that depends on it: if a commit is believed to be vulnerable, then all functions changed by
the commit are labeled vulnerable, ``which is always not true''~\cite{pseedcvefixesautomatedcollection}.
Its own release---5{,}365 CVE records across 1{,}754 projects in 5{,}495 fixing
commits---inherits that flaw wholesale, as does every corpus built the same way.

Two results establish how deep the problem goes. PatchDB reports the cost side: manual verification of
promising candidates confirms a security patch only 25\% of the time, so honest labeling is expensive
enough that its authors resort to synthesizing control-flow variants to populate the long
tail~\cite{pseedpatchdblargescale}, importing the synthetic-realism problem into the patch domain.
Mono reports the depth side, and it is the sharpest label finding in the corpus: alongside the 31.0\% of labels it corrects outright, a further \textbf{one CVE in six is \emph{undecidable}} from the code alone in the patch-derived benchmark it evaluates~\cite{p951}. Those cases are not determinable from the supplied static code alone, so they cannot be reliably learned under that representation without additional context; a model may instead fit correlates of the label.

Several downstream findings are consistent with this explanation. VulRepair achieves
perfect repair on some weakness types and zero on others, a profile that tracks CWE coverage in the
fix corpus rather than intrinsic difficulty~\cite{pseedvulrepairt5based}. SPVR and its baseline
together leave 355 samples that neither can repair, a residue larger than either method's unique
contribution~\cite{p195}. A Java benchmark of 2{,}362 validated vulnerabilities across 32 CWEs finds
patches that compile and score well on BLEU while failing to fix the vulnerability or introducing new
insecure patterns~\cite{p669}, similarity is not a security oracle. And the distribution itself is
skewed: detectors trained on NVD-linked patches lose up to 90\% F1 on in-the-wild security patches,
because the two populations differ enough to be told apart trivially~\cite{p220}.

The most consequential recent result pushes past labeling entirely. Testing code agents on real
repositories, Peng et al.\ show that 4.3--6.0\% of \emph{functionally correct} patches are vulnerable
with no attacker present, and that appending a developer-style suggestion to an issue description
drives that to 40.7\%~\cite{p768}. Passing the tests is the only criterion agents are held to, and it
is not a security oracle.

Beyond the deep-coded set, patch data is consumed overwhelmingly for repair, and the wider
corpus is dominated by LLM-based repair evaluated on mined fix pairs~\cite{bnguyen2025vulnerabil,balrashedy2025leveraging,bkado2025automated,bpearce2022code,bzhang2024trained,bgao2024sguard,p1578,bkhan2025code,brajput2022icspatch,bhou2025linejlocre,p1567,p1227}.

What is changing is the role of the patch. Rather than serving as a label, a patch is increasingly the
\emph{input} to exploit generation---turned into a working PoC for Linux-kernel
N-days~\cite{p284}, for Ethereum clients~\cite{p2111}, and for JavaScript packages where no
vulnerability report exists yet~\cite{p2110}. VulnResolver takes the same turn defensively, deriving
safety properties from code and validating them by PoC re-execution rather than relying on
annotations~\cite{p38}. That is the T3 row migrating toward T4, and it is the most promising movement
in the corpus.

\subsubsection{Tests, PoCs, and Exploits (T4)}\label{res:t4}
Executable artifacts sit at the top of the ladder, and among the types with enough datasets to carry a rate they are the only one for which majorities meet both the realism (16 of 24) and label-evidence (23 of 24) criteria in Figure~\ref{fig:matrix}. The reason is a
single structural property, named precisely by PBFuzz as the \emph{asymmetry of verification}: a
proof of vulnerability is hard to generate but easy to check~\cite{p2109}. An exploit that runs is therefore independently checkable evidence that the behaviour is reachable in the tested environment. Execution removes one major annotation dependency by making the triggered behaviour directly observable, which is why this row is less exposed to the labeling problems that shape the rows above it.

The same property that makes these artifacts directly checkable makes them scarce, and the corpus
quantifies the funnel at three separate points. Building VulnRepairEval required processing roughly
2{,}500 sources and 448 CVEs across 164 projects to yield \textbf{23} Python CVEs with working PoCs
amenable to automated testing---about 5\%~\cite{p778}. Magneto could include only 6 of 22 and 14 of 30
vulnerabilities from two prior benchmarks, excluding the rest because its authors could not manually
exploit them~\cite{p1277}. SEC-bench Pro must defer any report lacking either a concrete PoC or a
linked fix, because its two-sided oracle needs both~\cite{p203}.

Two studies measure the ceiling this implies. A large-scale study of PoCs in the wild finds that
\textbf{78.9\% of CVEs have no publicly available proof-of-concept}, and that the reports which do exist
omit roughly 30\% of the components needed to reproduce the flaw~\cite{p898}. A study of patching
practice across 26{,}803 CVEs locates the same gap one step earlier: \textbf{70.2\% of CVE patches
contain no test files}~\cite{p2110}. That study found no publicly available proof-of-concept for roughly four CVEs in five, and the artifact that would supply it is usually absent from the fix itself.

The size distribution follows directly. Executable benchmarks stay in the hundreds---118 instrumented
bugs in Magma~\cite{pseedmagmagroundtruth}, 98 scenarios in SecCodeBench-V2~\cite{p170}, 183 in
SEC-bench Pro~\cite{p203}, 318 tasks in SecRepoBench~\cite{p863}, 25 in ChainDelta~\cite{p2111}, 23 in
VulnRepairEval~\cite{p778}---while label-only corpora reach hundreds of thousands. The exception is informative: CyberGym reaches 1{,}507 vulnerabilities across 188 projects precisely because
OSS-Fuzz had already produced the ground-truth PoCs~\cite{p1109}.

Automating that production is the field's live attempt to escape the constraint, and it is advancing
quickly without yet closing the gap. Magma sidesteps generation by forward-porting real bugs into real
targets so that triggering is directly observable~\cite{pseedmagmagroundtruth}, \emph{yet} that works
only for bugs already understood. K-REPRO reproduces over half of 100 Linux-kernel N-days from their
patches~\cite{p284}, \emph{yet} temporal memory violations remain the residual hard case. PBFuzz
triggers 57 of 129 Magma CVEs, 17 of them unreached by any prior approach~\cite{p2109}, \emph{yet}
72 remain untriggered on a benchmark whose ground truth is fully known. ChainDelta generates exploits
for 16 of 25 Ethereum vulnerabilities and finds four undisclosed ones~\cite{p2111}, \emph{yet} nine
patched vulnerabilities resist it. In every case the automated fraction is real and the residue is the
hard part.

Two limits deserve statement because the benchmarks state them about themselves. SecCodeBench-V2 puts
the soundness--completeness distinction most sharply: execution-driven validation is an objective
oracle, but passing its tests ``does not necessarily imply the absence of vulnerabilities,'' since
attack paths not exercised by the PoCs remain unexplored---and a whole class of issues, including weak
cryptography, hardcoded credentials, and information leakage, cannot be adjudicated by tests at all
and falls back to an LLM-as-judge~\cite{p170}. FirmAgent states the complementary dichotomy: static
analysis, LLMs included, yields high false positives and no PoC, while fuzzing yields high false
negatives~\cite{p505}. Executable evidence is the strongest available, not a complete one.

Executable work is the second-largest group in the screened corpus, and the recurring pattern
across it is a benchmark of tens to a few hundred instances built by reconstructing
environments by hand~\cite{bsheng2026fuzzingbra,bli2026execution,bblain2026broken,bandroutsopoulo2025deepsurf,byan2025guiding,bxi2025trace,bnitin2025faultline,bsheng2025need,bliu2025quot,bwang2025aeppollo,bwen2025instructio,bsun2026viper,bzhang2026validating,bzhao2026triggering,bxie2026shadowprob,bliu2026scaling}.

Two further observations follow from the coding. First, contamination control concentrates
here---four of the corpus's contamination-controlled values are T4
benchmarks~\cite{p163,p170,p2109,p863}---because when the artifact is executable, memorization can be
measured and designed against rather than assumed away. Second, and against expectation, T4 shares the lowest availability rate among the major types: 12 of 24 are public, with releases promised on
acceptance~\cite{pseedtoolemu} or on publication~\cite{p778}. The artifacts that cost the most to build are among the least often shared---level with patch datasets at 50\%---which converts a scarcity problem into a duplication problem.

\subsubsection{Traces and Logs (T6)}\label{res:t6}
Execution traces, crash logs, and agent trajectories record what a system actually \emph{did}. Three deep-coded datasets carry traces as their primary artifact: a hand-labeled corpus of insecure agent trajectories~\cite{p766}, a released corpus of real coding-agent sessions~\cite{bbaumann2026chat}, and a gated release of multi-turn adversarial conversations with full chain-of-thought traces~\cite{p14}. A fourth contributes traces behind an executable safety benchmark,
where our tie-break rule (Section~\ref{sec:aspect-a}) makes the harness the primary type and records
traces as secondary~\cite{bhu2026saber}. Two further datasets carry execution records the same way---ToolEmu's LM-emulated runs~\cite{pseedtoolemu} and PBFuzz's fuzzing feedback~\cite{p2109}. Because the deep-coded row is small and the corpus now carries 94 trace-tagged
papers, we checked all 94 against their titles, abstracts, and screening records for a reusable trace contribution. \textbf{Six appeared to release a trace corpus as their primary artifact, five to release traces behind a benchmark or harness, and the remaining 83 to report trace-based techniques, attacks, or measurements with no reusable data artifact.} We then read the four candidates outside the anchor set in full, and the screen proved optimistic: one is a primary trace corpus and is now deep-coded~\cite{p14}, while three release a benchmark, a simulation venue, or a bargaining protocol whose runs emit traces, making the trace an evaluation record rather than the artifact. \textbf{Three primary trace corpora survive full-text assessment}, all from 2025 or later. Trace-related work is no longer rare; a released trace corpus is, and a screening-level reading of titles and abstracts systematically over-counts it---which is the more useful finding, since it is the reading most consumers of this literature will do.

The earliest of the three establishes why the row matters. Analyzing over 12{,}000 agent actions across five
models on 93 real-world software-setup tasks spanning seven languages, it finds that \textbf{21\% of
agent trajectories contain insecure actions during ordinary operation, with no attacker
involved}~\cite{p766}. The dominant weakness is CWE-200 (exposure of sensitive information), followed
by improper access control and downloading code without an integrity check. None of these is a
property of a diff. They live in the action stream---which file was read, which endpoint was called,
which artifact was fetched unverified---and are therefore invisible to every code-centric dataset in
this study, however large.

The cost of entry explains the scarcity. Before any detector could be built, the authors had to hand-label
500 trajectory steps as secure or insecure, because no labeled corpus of agent behavior
existed~\cite{p766}. Traces are cheap to \emph{produce} and expensive to \emph{label}, the inverse of
the code-sample economics, and unlike exploits they carry no self-certifying property: a trajectory
does not announce that it was unsafe.

The second dataset is recent and much larger: 6{,}000 coding-agent sessions logged from
open-source developers who opted in, comprising more than 63{,}000 user prompts and 355{,}000 tool
calls~\cite{bbaumann2026chat}. It shows the row is beginning to fill, and it inherits the labeling
economics above---the sessions are released, but their security properties are annotated by model
judges rather than by hand. Adjacent work on agent security is plentiful, but it attacks or defends
agents rather than releasing records of what they did.

This row matters increasingly as assistants become agents that take actions rather than suggest text, because the diff stops being the unit at which security can be assessed,
and the trace becomes it. Against that shift the deep-coded row holds three corpora, one of them hand-labeled at 500 steps---the total after the corpus-wide screen above and a full-text check of every candidate it raised.

\subsection{Study result: data quality (RQ2)---the core}\label{survey-result-data-quality-rq2-the-core}

We report each attribute two ways: how many papers \textbf{study} it, produce a finding about it, and how well the 93 dataset-contributing papers \textbf{achieve} it
(Table~\ref{tab:t64b_studies_vs_datasets}); Table~\ref{tab:t64_dist} gives the full value
distributions behind both columns. The gap between those two columns is this study's central
observation. Among the 111 deep-coded anchor papers, no attribute is the recorded object of a quality finding in more than fourteen, and the attribute
that most directly governs whether a reported score means anything, leakage, goes unaddressed by 49 of the 90 datasets where it applies. Realism, label evidence, and availability are almost always
assessable; what varies is how well they are met, not whether the dataset engages them at all.

The study counts themselves are informative: label evidence (14), leakage (8), realism (8), diversity (7), availability (6)---and \textbf{scale (0)}. Not one paper in the Tier-1 set produces a
finding about scale as such. We argue in Section~\ref{scale} that this is correct rather than
negligent: across the datasets we coded, achievable scale appears strongly constrained by the cost and coverage of the chosen labeling procedure. That relationship, quality attributes constraining one another rather than varying freely, is the thread running through the six subsections below.

\subsubsection{Realism}\label{realism}

Realism is \textbf{bimodal by artifact type}. Across the 93 instances the values split mined/semi-real 35, real-world 25, synthetic 18, injected/generated 14---but that aggregate hides the finding: the
25 real-world datasets are still mostly executable artifacts (T4: 16 of 24), whereas \textbf{just one of the 41 code-sample (T1) datasets is real-world} (Figure~\ref{fig:matrix}). The code the field
trains detectors on is synthetic, injected, or mined-and-simplified; authentically real-world data is concentrated in exploit and test data---16 of the 25 real-world datasets are T4, the other 9 spread across the remaining types, which is scarce (Section~\ref{availability}).

\textbf{What the composite hides.} Our
\emph{mined/semi-real} value is defined as material drawn from real projects or CVEs that is either
sliced to a function or file \emph{or} carries imperfect labels. Both clauses refer to properties we
code separately---granularity in Aspect~A, label evidence in Aspect~B---so a function-granularity
dataset cannot reach \emph{real-world} however authentic its vulnerabilities are. The figure above is
therefore not a free-standing discovery: it partly restates the coding rule. What the rule encodes is
an argument we defend in Section~\ref{res:t1}, that slicing to the function is where realism is lost, but the argument should be visible rather than buried in a value definition.

The components are separately observable, and reported apart they say more. Of the 41 code-sample
datasets, \textbf{20 draw their vulnerabilities from real projects or CVEs}, but only \textbf{3 keep
the sample at the unit the code is deployed in}, and \textbf{2} do both. Executable artifacts show the opposite profile on the same
two questions: 16 of 24 are graded real-world and 23 of 24 carry labels that received an independent check, with 15 meeting
both. That
contrast is the evidence-ladder finding stated without the composite: code-sample datasets are not
short of authentic vulnerabilities, half of them have one, they are short of the context in which
those vulnerabilities are reachable. We report the decomposition for code samples because that is where origin and context come
apart; \emph{context retained} here means the sample is shipped at module or repository granularity,
which is a coarser test than the codebook's full-context criterion and is not a substitute for the
realism grade.

Treating realism as one axis, then, obscures that the corpus fails it in three separable ways, and
a dataset can be strong on one while failing another. The first is \emph{provenance}: whether the
vulnerability was written by a human in real software, injected by a tool, or generated wholesale.
The second is \emph{context}: whether the sample retains the calling context, build configuration,
and inter-procedural flow in which the vulnerability manifests, or has been sliced to a function. The
third is \emph{distribution}: whether the ratio of vulnerable to benign code resembles deployment.
Our coding captures the first directly; the second appears as granularity, where the function dominates (50 of 111); the third surfaces only in the few papers that report base rates.

Separating them explains results that otherwise look contradictory. A dataset can be perfectly
authentic in provenance and still fail badly, because it was sliced. Big-Vul draws every sample from
real CVEs in real projects~\cite{pseedbigvul}, yet models trained on such corpora lose more than half
their performance in a realistic setting~\cite{pseedreveal}. The provenance was never the problem;
the context and the distribution were.

The eight realism studies converge on a single measurement: performance on unrealistic data does not
transfer, and the drop is larger in the cited studies when more of the three dimensions differ. Neural
code editors reach \textbf{79.40\%} accuracy on synthetic code and \textbf{10.03\%} on real-world
code, a collapse across provenance alone~\cite{pseedgeneratingrealisticvulnerabiliti}. ReVeal traces
its greater-than-50\% drop to context and distribution together~\cite{pseedreveal}.
Patch-identification models trained on NVD-linked data lose \textbf{up to 90\% F1} in the wild, where
the two populations differ enough to be told apart trivially---a pure distribution
failure with authentic provenance on both sides~\cite{p220}.

For AI-generated code the same decomposition applies with the terms shifted. Nearly all security
evidence about generated code comes from controlled prompt scenarios, which is a provenance
restriction of a new kind: the code is real output from real models, but elicited under conditions
that do not resemble use. Measured instead \emph{in the wild}, 29.5\% of Python and 24.2\% of
JavaScript AI-generated snippets already merged into GitHub carry weaknesses spanning 43 CWE
categories~\cite{p850}. The T1 row of Figure~\ref{fig:matrix} will not move until generation is
certified by an executable oracle, or until mining preserves repository-level context instead of
slicing it away.

\subsubsection{Label evidence}\label{label-correctness}

At first glance labels look healthy---\textbf{51 of 88} engaged datasets are \emph{verified}---but
the aggregate is an artifact of which types dominate. Conditional on a working harness, rechecking is comparatively inexpensive for executable artifacts, where a successful run supplies the check (T4: 23 of 24), and expensive everywhere else (T1: 14 of 36). Among mined code and patch datasets the dominant value is \textbf{patch-derived (24)}, with \textbf{tool-derived (11)} and \textbf{description-derived (2)} labels carrying a scanner's or a
CVE text's error rate directly into the ground truth.

What organizes this row is a cost hierarchy of label oracles. Execution is cheap to apply and hard to argue with, but available only where an exploit exists. Expert review provides stronger independent evidence, and is expensive: building Devign's corpus
took four professional security researchers 600 man-hours~\cite{pseeddevigneffectivevulnerability},
and PatchDB reports that manual verification confirms a security patch only 25\% of the
time~\cite{pseedpatchdblargescale}. The patch heuristic is cheap and wrong in a specific way, stated
by CVEfixes from inside the practice that relies on it: when a commit is deemed vulnerable, every
function it touched is labeled vulnerable, ``which is always not
true''~\cite{pseedcvefixesautomatedcollection}. Static analysis is cheaper still and inherits the
scanner's false positives~\cite{p826,p974}. CVE text is cheapest and worst, because CWE categories
overlap semantically and vary by language~\cite{p801}. The coded cases are consistent with verification cost constraining which evidential basis a dataset can use at scale.

Label evidence is the most-studied attribute (14 papers), and the studies are damning. A
label-error study finds errors \emph{endemic} to the datasets the field trains on, costing up to
\textbf{20.7\% F1} and recovering \textbf{10.4\%} on average under denoising---showing that label errors can materially distort reported performance~\cite{p1420}. Mono goes further: beyond
ordinary mislabeling, which it corrects at a rate of 31.0\%, \textbf{16.7\% of the CVEs it evaluates carry ``undecidable'' patches} whose root cause cannot be recovered from the supplied static context~\cite{p951}.
That distinction matters more than the percentages. Noisy data can be denoised; undecidable data cannot, because the intended label cannot be reliably resolved from the provided static-code context alone.
PrimeVul names the resulting syndrome outright---poor data quality, low label accuracy, high
duplication---and adds that the standard metrics misreport it~\cite{pseedprimevul}.

One case shows the hierarchy from the other end. Package hallucination admits a cheap \emph{sound}
oracle: a package either exists in a registry or it does not. With that oracle available, a single
study reports exact counts at \textbf{576,000-sample} scale, identifying 205,474 unique hallucinated
package names~\cite{p1222}. The contrast with the 600 man-hours behind Devign's few thousand samples
is the whole argument of this subsection. \textbf{For narrowly defined properties with a cheap and decisive oracle the binding constraint is economic}---wherever such an oracle exists, the field already produces independently checked labels at scale. For general software vulnerabilities the constraint is not cost alone: as Section~\ref{lim:oracle} sets out, labeling is also bounded by what can be decided from the artifact at hand and by what any single oracle covers.

\subsubsection{Scale}\label{scale}

Scale is the only attribute no paper in the Tier-1 set takes as its primary object. One result
bears on it indirectly---augmentation gains plateau past roughly 5k generated
samples~\cite{pseedvulscriberexploringrag}---but that is a finding about the returns to scale under a
fixed oracle, not about scale as a property to be characterised. We read the absence as a correct intuition rather than an oversight: across the datasets we coded, \textbf{achievable scale appears strongly constrained by the cost and coverage of the chosen label oracle}. Ranked by the cost of its label oracle, the corpus orders closely by size.

At the expensive end sit hand-verified datasets. SVEN is manually curated at 1,606 programs spanning
nine CWEs~\cite{pseedsven}; Devign labels manually across four
projects~\cite{pseeddevigneffectivevulnerability}. In the middle sit the patch-derived workhorses,
which buy an order of magnitude by accepting the tangled-commit error: Big-Vul reaches 3,754
vulnerabilities across 348 projects and 91 CWE types~\cite{pseedbigvul}, PrimeVul 6,968 vulnerable
against 228,800 benign functions across 755 projects~\cite{pseedprimevul}. At the cheap end sit
generators, where the oracle is the generation procedure itself: VinJ produces 686k samples at 0.4
seconds each, and 69\% of them are actually vulnerable, so roughly a third of the corpus is
mislabeled by construction~\cite{p1230}. And at the opposite extreme sits executable evidence, where
the oracle is sound but the artifact is nearly unobtainable: after processing approximately 2,500
sources and 448 CVEs, VulnRepairEval retained \textbf{23} instances with working
exploits~\cite{p778}.

Three orders of magnitude separate VinJ from VulnRepairEval, and the ordering runs broadly counter to label quality. This is why scale reported alone is uninformative, and why we record it as a
figure with its unit rather than as a bucket: 686k mislabeled samples and 23 exploit-verified ones are
not two points on one axis.

Scale can also show diminishing returns that are not about cost. In one augmentation setting, downstream gains plateaued past roughly 5k generated samples, after which additional synthetic data stopped helping and could reverse~\cite{pseedvulscriberexploringrag}. The mechanism generalises further than the threshold does: data drawn from the same distribution under the same oracle carries little new information whatever the artifact type. That is the case for reporting a scaling curve rather than assuming volume helps, and it is why the field's answer to its data problem is unlikely to be volume alone.

\subsubsection{Diversity}\label{diversity}

Diversity is reported on four facets---language, CWE coverage, project coverage, and class balance---and is uneven on all four. Seven papers study it, and, tellingly, most do so as a secondary
finding of a realism or leakage study. Diversity is widely observed and rarely measured on purpose.

\begin{figure}[t]
\centering
\includegraphics[width=\linewidth]{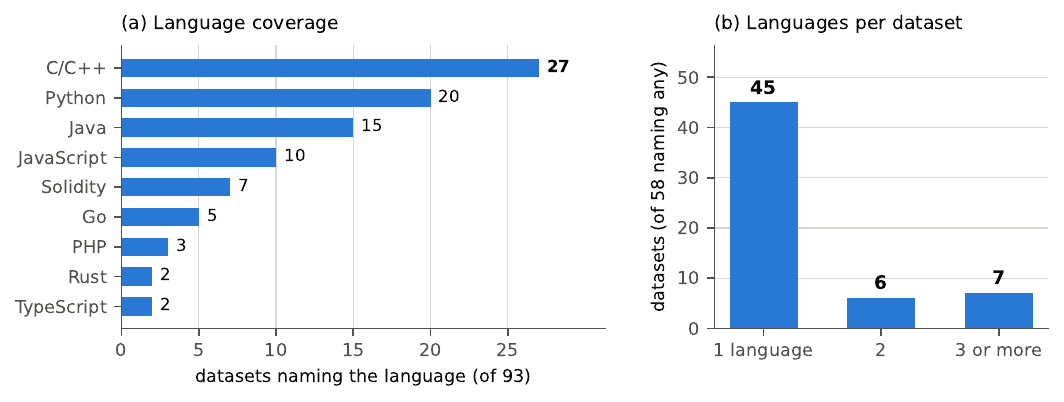}
\caption{Language coverage across the 93 contributed datasets. (a) Datasets naming each language;
33 name none, chiefly agent benchmarks and cross-dataset studies. (b) Languages per dataset, over the
58 that name any: coverage is not merely C/C++-weighted but siloed, with 45 datasets confined to a
single language and only 7 spanning three or more.}
\label{fig:language}
\end{figure}

\textbf{Language coverage.} Across the 93 contributed datasets, C/C++ appears in 27 (30\%), Python in 20, Java in 15, and JavaScript in 10,
with Solidity in 7 and Go in 5, and PHP, Rust, and TypeScript in three or fewer each. More revealing
than the ranking is the breadth: \textbf{45 of the 93 datasets are single-language, and only 7 cover three or more}. Twenty-one name C/C++ and nothing else. The field does not have a multilingual data problem so much as a
collection of single-language silos, of which C/C++ is the largest group (Figure~\ref{fig:language}).

The consequences show up wherever a second language is attempted. Java was served only recently, by a
benchmark of 30,637 methods spanning 1,740 CVEs, 700+ projects, and 225 CWE
categories~\cite{p412}, and by a repair benchmark of 2,362 validated vulnerabilities across 20
projects and 32 CWEs whose authors note that the prior standard, Vul4J, offered 79
vulnerabilities~\cite{p669}. Other ecosystems remain served by isolated one-off efforts: smart
contracts~\cite{p974,p287}, the Ethereum client stack~\cite{p2111}, IoT firmware~\cite{p505}, and the
JavaScript package ecosystem~\cite{p2110}. Each is a first dataset in its domain, which is a measure
of how thin coverage is outside C/C++.

\textbf{Class balance} is a second systematic problem, and it is the facet whose violation is least often reported. Real vulnerable code is rare---PrimeVul's mined corpus runs roughly 1 vulnerable function
to 33 benign~\cite{pseedprimevul}---so a research dataset that balances its classes misrepresents
the deployment base rate by more than an order of magnitude. PrimeVul argues this invalidates accuracy
and F1 as headline metrics outright, and an open-science replication finds that balanced or
artificially generated datasets produced significantly overrated performance for the techniques
evaluated on them~\cite{p1433}.

\textbf{CWE coverage} is long-tailed everywhere it is reported. PatchDB's NVD-derived component is
explicitly long-tailed and highly imbalanced, which is why its authors resort to synthesizing
control-flow variants to populate the tail~\cite{pseedpatchdblargescale}. The downstream signature is
visible in repair: VulRepair achieves perfect prediction on some weakness types and zero on
others, a profile that tracks corpus coverage rather than intrinsic
difficulty~\cite{pseedvulrepairt5based}, while SPVR caps its scope at the 2023 CWE Top-25 by
design~\cite{p195}. Reported repair and detection rates are therefore partly reports about which CWEs
happen to be well represented.

\subsubsection{Leakage and contamination}\label{leakage-and-contamination}

Leakage is where the datasets we coded are weakest, by the plainest measure: \textbf{49 of the 90 datasets where it applies (55\%) do not address train/test independence at all}. Among the 41 that do, the strategies are deduplication (15), chronological (13), contamination-controlled (5), random-with-none (5), and project-disjoint (3).

That count is generous, and the generosity is worth making explicit. Recording a split strategy is not
the same as guarding against leakage: a random split with no deduplication is the \emph{highest}-risk
configuration, since it neither detects nor separates duplicates. Counting those five with the non-reporting datasets, \textbf{54 of 90 do not report an effective leakage guard}. The two figures answer different questions and we keep them distinct throughout: \emph{49 of 90} is how many say nothing about leakage at all, and \emph{54 of 90} is how many report no effective guard, the difference being the five that report a random split without deduplication. Among the datasets we deep-code, JavaVulBench is the only one that provides several split strategies together (random, project-disjoint, temporal, deduplicated, and unseen-CWE-family) with a per-model contamination audit~\cite{p412}.

Eight papers study leakage, and they show why this omission matters. ReVeal traces its
greater-than-50\% real-world performance drop partly to \textbf{up to 68\% duplication} between train
and test~\cite{pseedreveal}; PrimeVul adopts chronological splitting specifically to reduce
it~\cite{pseedprimevul}. For LLMs, ordinary train/test leakage is joined by \textbf{pretraining contamination}, which requires different controls. Models score high on public benchmarks their training data has seen and then collapse on post-cutoff zero-day sets~\cite{pseedassessingimprovingprompting}; Copilot reproduces original vulnerable code roughly 33\% of the time, and more often for \emph{older} vulnerabilities~\cite{p1494}. Because training corpora are typically undisclosed, contamination often has to be assessed indirectly; perturbation sensitivity across benchmarks is one such diagnostic~\cite{p378}. Recent datasets increasingly control contamination by \emph{construction} rather than only through data splitting: four of the five contamination-controlled datasets in the corpus
are executable benchmarks. SecCodeBench-V2 is curated from de-identified proprietary industrial
cases that cannot overlap public training corpora~\cite{p170}; SecRepoBench applies semantic-preserving
mutation to defeat memorization~\cite{p863}; PBFuzz excludes CVEs with public ready-to-run
proofs-of-vulnerability~\cite{p2109}; SecureVibeBench selects tasks to avoid vulnerability-introducing
contexts~\cite{p163}. The pattern is not coincidental: when an artifact is executable, memorization
can be measured and designed against, whereas for a static label contamination can only be hoped
away. The fifth case shows the practice starting to spread beyond that setting---SecPI, a
code-sample dataset, argues instead from model knowledge cutoffs predating its evaluation benchmarks
and evaluates on held-out CWE families~\cite{p168}. That argument is weaker than a mutation or a
proprietary corpus, since it rests on a claim about the model rather than a property of the data, but
it is the first instance in the corpus of a non-executable dataset attempting the guard at all.

One asymmetry deserves emphasis. Unlike correct labels, which cost hundreds of expert hours, a
chronological or project-disjoint split needs little beyond reliable timestamps or project
identifiers and a decision about how to partition data one already holds. It is not free---a temporal
split can shrink the training set, shift the class balance, and make evaluation harder---but the cost
falls once, in implementation, rather than per sample. That 55\% do not report addressing it suggests that reporting and evaluation norms contribute materially to the leakage gap, especially because documenting a split requires no per-sample labeling effort.

\subsubsection{Availability}\label{availability}

Availability is stated more often than not, but non-reporting remains substantial and one-sided:
\textbf{25 of 93 datasets make no availability statement at all, 62 are public, and 5 are
explicitly withheld}---the latter typically promising
release ``upon acceptance''~\cite{pseedtoolemu} or ``upon publication''~\cite{p778}. (The
not-reported figure is a lower bound; see the Section~\ref{threats-to-validity} extraction caveat.)
That 25 papers describe a dataset without saying whether anyone can obtain it is itself the
finding:
availability is not yet treated as a property worth reporting.

Availability is lowest for some of the artifacts that are most expensive to reconstruct. Executable artifacts and patch datasets have the lowest public-availability rates, at 12 of 24 and 7 of 14 respectively, compared with 33 of 41 code-sample datasets (Figure~\ref{fig:matrix}). Building a repository-level exploit harness can require months of environment reconstruction~\cite{p203}, so withholding such an artifact forces later researchers to repeat substantial engineering work.

Availability problems also accumulate across the reproducibility process. In the study that measured the full process, only 25.5\% of reviewed deep-learning vulnerability detectors provided a public tool; among the reproducible tools, only 14.3\% were partially and \textbf{28.6\% fully replicable} on another real-world dataset~\cite{p1433}. At a more basic level, some evidence is unavailable because it was never produced: most disclosed vulnerabilities have no publicly available proof-of-concept~\cite{p898}, and most patches ship without a test that demonstrates the vulnerability~\cite{p2110}. Availability therefore depends not only on whether existing artifacts are released, but also on whether strong evidence was produced in the first place. And a systematic
review states the resulting oracle problem directly: exploits are what repair techniques need in order
to verify a patch, yet they are usually unavailable and require security expertise to
craft~\cite{p1201}. Together, these results show that availability is limited both by release practices and by the cost of producing strong evidence in the first place.

\subsubsection{The attributes are not independent}\label{attributes-not-independent}

Read together, the six attributes form a dependency structure rather than a checklist, and that structure accounts for much of the variation we observe across the coded datasets.

Labeling strategy links several of these attributes. What a dataset can afford to verify is closely tied to how large it grows
(Section~\ref{scale}), and the granularity it must slice to in order to make automatic labeling
tractable is the same choice that costs it context, and therefore realism
(Section~\ref{realism}). Both together appear to constrain diversity: a corpus that can only afford patch-derived
labels in C/C++ will cover the CWEs that appear in C/C++ fix commits, and no others
(Section~\ref{diversity}). Availability then determines whether any of this is reusable, and for
executable artifacts, the only ones that mitigate a major part of the oracle problem, it is mostly not
(Section~\ref{availability}).

Leakage stands outside this structure, which is why it is the most tractable of the six. It adds no per-sample labeling effort, depends on no oracle, and is, for ordinary train/test leakage, largely a matter of how one partitions data already in hand; guarding against pretraining contamination can additionally require dataset construction or model-specific auditing.
That it remains unaddressed in a majority of the applicable datasets we coded suggests that convention is an important constraint alongside cost, and conventions are the part a study can hope to change.

\subsection{Study result: consumption and evaluation (RQ3)}\label{survey-result-consumption-and-evaluation-rq3}

\textbf{Tasks.} The consumption picture below characterises the \emph{anchor set}, not the field: task and technique are hand-coded and therefore available only for the 111 deep-coded papers, and that set is purposive by construction (Section~\ref{threats-to-validity}), admitting emerging agent, reasoning, and trace work deliberately. Read the figures as what the field's anchor artifacts do, not as shares of the literature. Within the anchor set, task centers on \textbf{detection (60)} and
\textbf{behavior/safety vetting (27)}---the latter a category that barely existed three years
ago, reflecting the shift toward evaluating generated code and agent actions---followed by
repair (25), validation (12), classification (11), localization (8) and explanation (5).
Most serve exactly one: 51 datasets are single-task, against 27 tagged with two and 7 with three.

The interesting structure is not the ranking but which artifacts serve which task, and how strong
that evidence is (Table~\ref{tab:t65_consumption}). \textbf{Validation is the only task supported
entirely by executable artifacts}---all 12 contributing datasets are T4, and 10 of them are
real-world---which is unsurprising, since validating a patch means running something. Every other
task rests substantially on weaker evidence. Detection, the largest, draws most of its records from code samples and only 25\% from executable
artifacts.

Three tasks stand out for having \emph{no} executable backing at all. Localization, classification
and explanation are served entirely by non-executable datasets, and between one and two datasets in each are
real-world. This is the
sharpest task-level asymmetry in the corpus, and it runs the wrong way: locating a flaw to a
statement and assigning it a CWE are the two tasks that most depend on the label being precisely
right, and they are supported by exactly the artifacts whose labels are least verifiable. A
patch-derived label tells you which lines a commit touched, not which line was vulnerable---the
tangled-commit problem of Section~\ref{label-correctness}---yet that is the evidence base
line-level localization is trained and evaluated on.

\textbf{Techniques.} By technique, \textbf{LLM prompting (51)} is more common than \textbf{deep learning (31)} among the anchors, with \textbf{benchmark-style evaluation (39)} and \textbf{agentic (24)} consumption concentrated in the recent ones. Split by era within this set: prompting appears in 2 of 17 pre-2023 records, 12 of 27 in 2023--24, and 37 of 67 in 2025--26; agentic consumption goes from none, to 3, to 21 over the same bands, while deep learning falls from 10 of 17 to 11 of 67.

One counter-current deserves note, because it contradicts the obvious reading of that trend.
Traditional program analysis remains broadly stable in proportional representation across the anchor cohorts, declining modestly from 24\% to 19\%: it appears in 4 of 17 early records, 6 of 27, and 13 of 67 recent ones. What
changes is its role. The recent instances are not holdouts but components---a
directed fuzzer inside an agentic loop~\cite{p2109}, fuzzing paired with an LLM agent over IoT
firmware~\cite{p505}, differential fuzzing driving exploit generation for Ethereum~\cite{p2111}. What the agentic era revived is precisely the capability LLMs lack: an execution-based oracle that can confirm whether an input actually triggers a fault, on the paths it exercises. The trajectory is therefore not a replacement sequence
but an accumulation, and it explains why the strongest-evidence artifacts (Section~\ref{res:t4})
cluster in the most recent period.

\textbf{Training versus evaluation.} The clearest structural finding in this section is a dichotomy
that runs along the artifact boundary. Of the 41 contributed code-sample datasets, 25 are consumed by a learned model through deep
learning or fine-tuning and just \emph{one} by an agentic pipeline. Of the 24 executable datasets,
\emph{none} is used for training at all; 17 are consumed as benchmarks and 16 agentically. Patch
datasets straddle the line, with 7 trained on and 3 consumed agentically.

Within the anchor set, the data is sharply partitioned: code samples are what systems learn from, executable
artifacts are what they are measured against, and almost nothing crosses over. That partition
explains a pattern this study returns to repeatedly. Improvements concentrated in executable
artifacts---their realism, their label evidence, their contamination controls---improve
\emph{measurement} without touching the substrate that models actually learn from, which is why the
aggregate quality trend of Section~\ref{sec:trajectory} rises while the code-sample cohort does not.
It also identifies the transfer that would matter most: making executable validation cheap enough to
apply to training data, not only to benchmarks, would be among the highest-leverage changes for letting the two halves of the corpus improve together.

\begin{figure}[t]
\centering
\includegraphics[width=\linewidth]{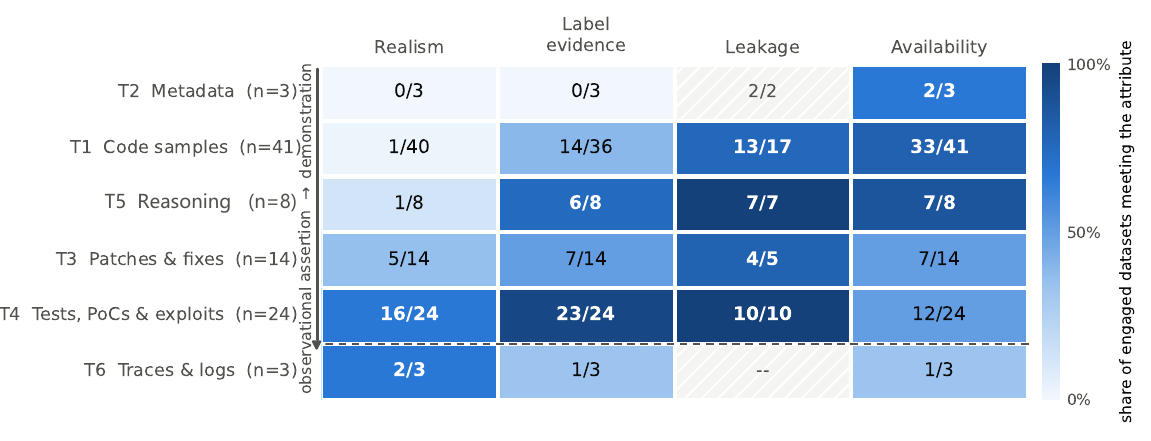}
\caption{Artifact $\times$ quality matrix over the 93 contributed datasets in the purposively selected anchor set. Each cell gives the number of datasets meeting the attribute over the number that engage it at all; shading is that fraction. The fractions characterise the anchor set rather than estimate field-wide prevalence, and part of the artifact-quality association follows from the taxonomy itself (Section~\ref{survey-result-data-quality-rq2-the-core}). Rows run with T2--T4 in demonstrative order and the reasoning and trace rows shown after them as explanatory and observational categories. Cells are hatched, and left unshaded, where fewer than
three datasets engage the attribute or fewer than 30\% of the row does---a fraction over one or two
datasets is not a magnitude, and shading it would let a single dataset read as maximal strength.
Part of what the matrix shows follows from the taxonomy rather than from independent measurement, and we separate the two. Executability is itself one of the routes by which our rubric recognises an independently checked label, so T4's 23 of 24 on label evidence is largely definitional: its significance is descriptive, that executable datasets carry the check with the artifact far more routinely than other forms do, not that artifact type causes verification. The realism grade likewise rewards retained deployment context, so a row's realism partly restates how its samples are cut. The contrasts that are \emph{not} implied by artifact type---leakage handling, availability, scale, diversity---carry the independent information, and they are where the row differences are least anticipated by the rubric. Read that way: realism strengthens down the ladder to the executable row, from 0\% at metadata to 67\% at tests and exploits; the trace row sits apart, as an observational record rather than a stronger rung, and rests on three deep-coded datasets. Label evidence strengthens with demonstrative strength overall but not step by step: it dips from 75\% at reasoning to 50\% at patches before reaching 96\% at executable artifacts. Availability moves the
other way.}
\label{fig:matrix}
\end{figure}
\subsection{Synthesis: the status quo at a glance}\label{synthesis-the-status-quo-at-a-glance}

The artifact × quality matrix (Figure~\ref{fig:matrix}) makes the pattern legible.
\textbf{Quality broadly tracks the evidence ladder}---as an association across the rows, not a
monotone law; label evidence dips at patches (Figure~\ref{fig:matrix}). Tests/PoCs/exploits (T4) are the only well-populated type for which majorities meet both the realism (16/24) and label-evidence (23/24) criteria---the two results have different causes: executability explains the label-evidence figure, since a run checks the triggered behaviour directly, while the realism figure reflects authentic targets with their deployment context retained. Code samples (T1), the largest group, are the
weakest on realism (one real-world instance in 41) and only partially verified (14/36).
Two failures then cut across every row: \textbf{49 of the 90 datasets where it applies never
address leakage}, and availability is public for 67\%.

The matrix also identifies two gaps that motivate the open problems in Section~\ref{sec:discussion}. First, \textbf{explanatory and observational artifacts remain sparsely represented}: traces (T6) rest on three deep-coded datasets, while reasoning (T5) has grown to eight since 2024. Reusable artifacts for explaining why code is considered vulnerable or recording what an agent did therefore remain limited in the deep-coded set, even as additional trace releases appear across the corpus. Second, \textbf{direct study of vulnerability-data quality remains limited}: no quality attribute has been the primary object of more than fourteen papers. Section~\ref{sec:discussion} develops these two gaps into research directions.

\subsection{Trajectory: is the picture improving?}\label{sec:trajectory}

Figure~\ref{fig:matrix} is cross-sectional. Because our records carry
publication years, we can also ask whether the picture is getting better, and the
answer depends entirely on whether one holds artifact type fixed.

\begin{figure}[t]
\centering
\includegraphics[width=\linewidth]{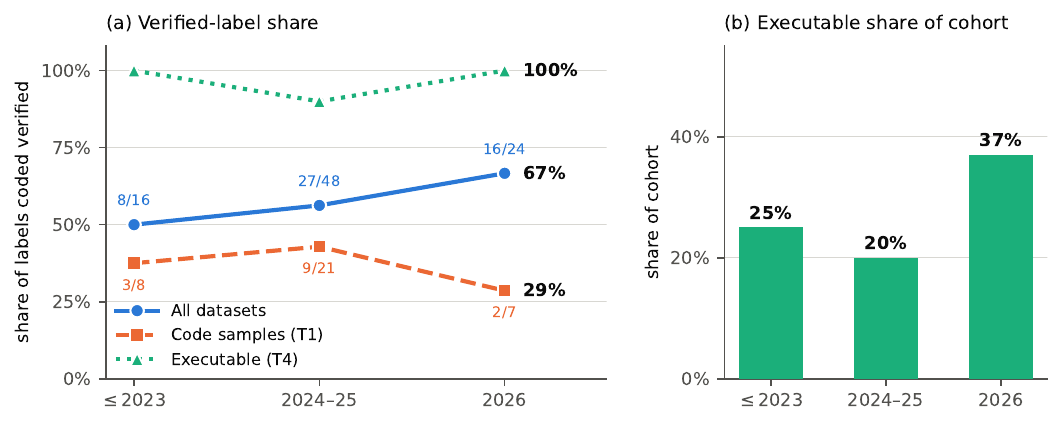}
\caption{Verified-label share over time. (a) The aggregate (blue) rises, but
held within artifact type it does not: code samples (orange) decline, while
executable artifacts (green) are near-total throughout because a successful run supplies the check directly. Fractions beside each point give the underlying counts; the aggregate denominators are the datasets whose label evidence was assessed, which is fewer than the band totals. (b) The aggregate rises because the cohort composition
shifts: executable artifacts grow from 25\% to 37\% of new datasets. Bands contain 16, 50, and 27 datasets respectively.}
\label{fig:trend}
\end{figure}

Among the datasets whose label evidence was assessed, it appears to improve
steadily: the share coded \emph{verified} rises from 50\% among datasets published through 2023 (8 of 16), to 56\% in 2024--25 (27 of 48), and stands at 67\% in 2026 (16 of 24)
(Figure~\ref{fig:trend}a, solid line). Held within artifact type, that trend
disappears. Among executable artifacts the verified share is flat and near-total
throughout, since a successful run supplies the check directly. Among code samples it moves
in the opposite direction, from 3 of 8 through 2023 to 2 of 7 in 2026. What
changed is the composition of the cohort: executable artifacts grew from 25\% of new datasets in the first period to 37\% in the last
(Figure~\ref{fig:trend}b). Within this set the aggregate improved because the more recent anchors are more often
executable, not because the code-sample anchors got better. Because the anchor set is purposively
selected, and selection deliberately admits rare and recent artifact types, we read this as a
pattern in the anchors rather than a population trend.

The same pattern holds for the other two attributes with a temporal signal.
Effective leakage handling rises and then falls back, and once executable
benchmarks are excluded the 2026 code-sample cohort sits close to where the
pre-2024 one did. Public availability has not recovered its earliest level.
Realism barely moves in the row where it is weakest: across all three periods exactly one code-sample dataset reaches real-world realism, so that finding is stable across the three anchor cohorts rather than a recent regression.

These bands contain 16, 50, and 27 datasets, and the 2026 code-sample cell rests on seven, so we read the trajectory as indicative. The direction is nonetheless consistent across label
evidence, leakage, and availability, and we take up what it implies in
Section~\ref{sec:conclusion}.

%% ------- auto-generated tables (scripts/export_tables.py maintains this block) -------
%%% BEGIN-AUTOTABLES
%% Auto-generated by scripts/export_tables.py — do not hand-edit inside this block.

\begin{table}[t]
\centering
\caption{Artifact-type distribution across the whole corpus (auto-tagged), the deep-coded Tier-1 set, and the datasets they contribute}
\label{tab:t61_type}
\small
\begin{tabular}{llrrr}
\toprule
Type & Name & Corpus & Tier-1 & Datasets \\
\midrule
T2 & Metadata & 119 & 5 & 3 \\
T1 & Code samples & 753 & 55 & 41 \\
T5 & Reasoning & 24 & 8 & 8 \\
T3 & Patches \& fixes & 156 & 15 & 14 \\
T4 & Tests, PoCs \& exploits & 376 & 25 & 24 \\
T6 & Traces \& logs & 94 & 3 & 3 \\
\midrule
 & Total & 1,522 & 111 & 93 \\
\bottomrule
\end{tabular}
\end{table}

\begin{table}[t]
\centering
\caption{Studies vs. datasets per attribute (93 dataset+both, 18 pure studies)}
\label{tab:t64b_studies_vs_datasets}
\small
\begin{tabular}{lrrr}
\toprule
Attribute & Papers that study it & Datasets strong / engaged & Datasets not engaging \\
\midrule
Realism & 8 & 25/92 & 1 \\
Label evidence & 14 & 51/88 & 5 \\
Leakage & 8 & 36/41 & 49 \\
Availability & 6 & 62/93 & 0 \\
\bottomrule
\end{tabular}
\end{table}

\begin{table}[t]
\centering
\caption{Value distributions for the four graded quality attributes (93 contributed datasets)}
\label{tab:t64_dist}
\small
\begin{tabular}{llr}
\toprule
Attribute & Value & Datasets \\
\midrule
Realism & mined/semi-real & 35 \\
 & real-world & 25 \\
 & synthetic & 18 \\
 & injected/generated & 14 \\
 & not assessed & 1 \\
\midrule
Label evidence & verified & 51 \\
 & patch-derived & 24 \\
 & tool-derived & 11 \\
 & not assessed & 5 \\
 & description-derived & 2 \\
\midrule
Leakage & not addressed & 49 \\
 & deduplicated & 15 \\
 & chronological & 13 \\
 & contamination-controlled & 5 \\
 & random (no dedup) & 5 \\
 & n/a (no learned consumer) & 3 \\
 & project-disjoint & 3 \\
\midrule
Availability & public & 62 \\
 & not reported & 25 \\
 & unavailable & 5 \\
 & on request & 1 \\
\bottomrule
\end{tabular}
\end{table}

\begin{table}[t]
\centering
\caption{What serves each consuming task: the datasets tagged with it, their dominant artifact type, and how many rest on executable or real-world evidence}
\label{tab:t65_consumption}
\small
\begin{tabular}{lrlrr}
\toprule
Consuming task & Datasets & Dominant type & Executable & Real-world \\
\midrule
Validation & 12 & T4 & 12 (100\%) & 10 (83\%) \\
Detection & 49 & T1 & 12 (24\%) & 12 (24\%) \\
Behavior/safety vetting & 26 & T1 & 12 (46\%) & 8 (31\%) \\
Repair & 24 & T3 & 3 (12\%) & 8 (33\%) \\
Localization & 8 & T1 & 0 (0\%) & 1 (12\%) \\
Classification & 11 & T1 & 0 (0\%) & 2 (18\%) \\
\bottomrule
\end{tabular}
\end{table}

\begin{longtable}{llllllll}
\caption{Paper attribution, part I (T2/T1/T5; 68 papers). Realism RW real-world, MS mined/semi-real, IG injected/generated, SY synthetic. Labels Ver verified, Pat patch-derived, Tool tool-derived, Desc description-derived. Leakage Chr chronological, Ded deduplicated, Prj project-disjoint, Con contamination-controlled, Rnd random, none not addressed. Availability Pub public, Req on request, No withheld. Kind D dataset, S study, B both. -- not assessed or not reported.}\label{tab:t5x_attribution_a} \\
\toprule
Paper & Yr & Gran & Real & Label & Leak & Avail & Kind \\
\midrule
\endfirsthead
\multicolumn{8}{l}{\small\itshape Paper attribution, part I (T2/T1/T5; 68 papers) (continued)} \\
\toprule
Paper & Yr & Gran & Real & Label & Leak & Avail & Kind \\
\midrule
\endhead
\midrule \multicolumn{8}{r}{\small\itshape continued on the next page} \\
\endfoot
\bottomrule
\endlastfoot
\textbf{T2 Metadata} &  &  &  &  &  &  &  \\
Nie et al. \cite{p1420} & 2023 & func & MS & Pat & Ded & Pub & B \\
Zhao et al. \cite{p801} & 2025 & func & SY & Desc & none & -- & D \\
ReVul-CoT \cite{p899} & 2025 & -- & MS & Desc & Prj & Pub & D \\
Young et al. \cite{p232} & 2026 & -- & -- & -- & none & Req & S \\
He et al. (Bridging) \cite{p636} & 2026 & -- & -- & -- & Ded & Pub & S \\
\midrule
\textbf{T1 Code samples} &  &  &  &  &  &  &  \\
Devign \cite{pseeddevigneffectivevulnerability} & 2019 & func & MS & Ver & Rnd & -- & D \\
Fan et al. \cite{pseedbigvul} & 2020 & func & MS & Pat & none & Pub & D \\
Nong et al. (Preliminary) \cite{pseedpreliminarystudyopen} & 2020 & file & SY & Ver & none & -- & S \\
Pearce et al. \cite{p1520} & 2021 & func & SY & Ver & none & Pub & D \\
D2A \cite{p1828} & 2021 & stmt & MS & Tool & Ded & Pub & D \\
Li et al. (Detection) \cite{p1870} & 2021 & func & MS & Pat & Rnd & Pub & S \\
Nong et al. (Evaluating) \cite{pseedevaluatingcomparingmemory} & 2021 & file & SY & Ver & none & -- & S \\
Chakraborty et al. \cite{pseedreveal} & 2021 & func & MS & Pat & Rnd & -- & S \\
VUDENC \cite{p1482} & 2022 & stmt & MS & Pat & none & Pub & D \\
Asare et al. (GitHub) \cite{p1494} & 2022 & func & MS & Pat & Chr & Pub & B \\
Schaad et al. \cite{p1499} & 2022 & func & SY & Ver & none & Pub & S \\
Nong et al. (Generating) \cite{pseedgeneratingrealisticvulnerabiliti} & 2022 & func & IG & Pat & none & -- & S \\
Nong et al. (Open) \cite{p1433} & 2023 & func & -- & -- & Rnd & -- & S \\
Steenhoek et al. (Detection) \cite{p1470} & 2023 & func & MS & Pat & Prj & Pub & S \\
He et al. (Security) \cite{pseedsven} & 2023 & func & MS & Ver & Rnd & Pub & D \\
VulGen \cite{pseedvulgenrealisticvulnerable} & 2023 & func & IG & Pat & Ded & -- & D \\
Spracklen et al. \cite{p1222} & 2024 & func & IG & Ver & Chr & No & B \\
VulnLLMEval \cite{p1227} & 2024 & mod & MS & Pat & none & No & B \\
VinJ \cite{p1230} & 2024 & func & IG & Pat & none & Pub & D \\
Tony et al. \cite{p1264} & 2024 & func & SY & -- & none & Pub & B \\
PromSec \cite{p1265} & 2024 & file & SY & Tool & none & Pub & B \\
Liu et al. \cite{p1276} & 2024 & func & MS & Ver & Rnd & Pub & D \\
Tóth et al. \cite{p1286} & 2024 & file & SY & Ver & none & Pub & B \\
Hamer et al. \cite{p1295} & 2024 & file & SY & Tool & none & Pub & B \\
Steenhoek et al. (Dataflow) \cite{p1356} & 2024 & func & MS & Pat & Prj & Pub & S \\
Asare et al. (User-centered) \cite{p1409} & 2024 & func & SY & Ver & none & Pub & B \\
Nguyen et al. \cite{p1624} & 2024 & stmt & MS & Pat & Chr & Pub & B \\
VGX \cite{pseedvgxlargescale} & 2024 & func & IG & Pat & Ded & -- & D \\
Cotroneo et al. \cite{p1021} & 2025 & func & MS & Tool & none & Pub & B \\
Mou et al. \cite{p1133} & 2025 & file & SY & Tool & none & Pub & B \\
Yildiz et al. \cite{p1153} & 2025 & func & MS & Pat & none & Pub & B \\
Das et al. \cite{p1614} & 2025 & func & SY & Ver & none & Pub & B \\
Risse et al. \cite{p807} & 2025 & func & MS & Ver & none & Pub & B \\
Dora et al. \cite{p818} & 2025 & mod & IG & Ver & none & -- & S \\
Xia et al. \cite{p826} & 2025 & func & MS & Tool & none & -- & D \\
Yu et al. (Smart-LLaMA-DPO) \cite{p843} & 2025 & mod & RW & Ver & none & Pub & B \\
Fu et al. \cite{p850} & 2025 & func & RW & Ver & none & Pub & S \\
Schreiber et al. \cite{p851} & 2025 & file & MS & Tool & none & -- & S \\
Secure-Instruct \cite{p857} & 2025 & func & IG & Tool & Ded & Pub & D \\
Dai et al. \cite{p888} & 2025 & func & SY & Tool & none & -- & S \\
Yu et al. (Unlocking) \cite{p976} & 2025 & func & MS & -- & Chr & -- & D \\
Gnieciak et al. \cite{p982} & 2025 & repo & IG & Ver & none & Pub & B \\
Ding et al. \cite{pseedprimevul} & 2025 & func & MS & Ver & Chr & Pub & B \\
VulScribeR \cite{pseedvulscriberexploringrag} & 2025 & func & IG & Pat & Prj & Pub & D \\
Kharma et al. \cite{p157} & 2026 & file & IG & Ver & none & Pub & B \\
SecureCodeRL \cite{p165} & 2026 & file & SY & Tool & none & Pub & S \\
SecPI \cite{p168} & 2026 & func & SY & Tool & Con & Pub & D \\
Tran et al. \cite{p274} & 2026 & func & IG & -- & none & Pub & D \\
ParaVul \cite{p287} & 2026 & file & MS & -- & none & -- & D \\
Jahromi et al. \cite{p313} & 2026 & func & SY & Ver & none & Pub & B \\
Euraste et al. \cite{p378} & 2026 & func & -- & -- & none & Pub & B \\
JavaVulBench \cite{p412} & 2026 & func & MS & Pat & Chr & Pub & D \\
Chen et al. \cite{p429} & 2026 & stmt & MS & Pat & Rnd & Pub & D \\
Nong et al. (Assessing) \cite{pseedassessingimprovingprompting} & 2026 & func & MS & Pat & Chr & Pub & B \\
Yi et al. \cite{pseedexploringimprovingreal} & 2026 & func & IG & Pat & none & Pub & B \\
\midrule
\textbf{T5 Reasoning} &  &  &  &  &  &  &  \\
Smart-LLaMA \cite{p1246} & 2024 & mod & MS & Ver & Ded & Pub & B \\
Lin et al. \cite{p1035} & 2025 & func & MS & Pat & Ded & Pub & D \\
Vul-R2 \cite{p1567} & 2025 & func & MS & Ver & Chr & Pub & B \\
VADER \cite{p791} & 2025 & file & RW & Ver & none & Pub & B \\
Yu et al. (Secure) \cite{p800} & 2025 & mod & MS & Ver & Ded & -- & D \\
Mao et al. \cite{p803} & 2025 & func & SY & Ver & Ded & Pub & B \\
R2Vul \cite{p907} & 2025 & func & MS & Pat & Ded & Pub & D \\
Li et al. (SFT) \cite{p486} & 2026 & mod & MS & Ver & Chr & Pub & B \\
\end{longtable}

\begin{longtable}{llllllll}
\caption{Paper attribution, part II (T3/T4/T6; 43 papers). Realism RW real-world, MS mined/semi-real, IG injected/generated, SY synthetic. Labels Ver verified, Pat patch-derived, Tool tool-derived, Desc description-derived. Leakage Chr chronological, Ded deduplicated, Prj project-disjoint, Con contamination-controlled, Rnd random, none not addressed. Availability Pub public, Req on request, No withheld. Kind D dataset, S study, B both. -- not assessed or not reported.}\label{tab:t5x_attribution_b} \\
\toprule
Paper & Yr & Gran & Real & Label & Leak & Avail & Kind \\
\midrule
\endfirsthead
\multicolumn{8}{l}{\small\itshape Paper attribution, part II (T3/T4/T6; 43 papers) (continued)} \\
\toprule
Paper & Yr & Gran & Real & Label & Leak & Avail & Kind \\
\midrule
\endhead
\midrule \multicolumn{8}{r}{\small\itshape continued on the next page} \\
\endfoot
\bottomrule
\endlastfoot
\textbf{T3 Patches \& fixes} &  &  &  &  &  &  &  \\
CVEfixes \cite{pseedcvefixesautomatedcollection} & 2021 & stmt & MS & Pat & none & Pub & D \\
PatchDB \cite{pseedpatchdblargescale} & 2021 & stmt & MS & Ver & Rnd & -- & D \\
VulRepair \cite{pseedvulrepairt5based} & 2022 & func & MS & Pat & none & -- & D \\
APPATCH \cite{p1403} & 2024 & func & RW & Ver & Chr & Pub & D \\
Le et al. \cite{p1410} & 2024 & stmt & SY & Ver & none & Pub & B \\
Kulsum et al. \cite{p1578} & 2024 & stmt & MS & Ver & none & Pub & S \\
SecureFixAgent \cite{p1565} & 2025 & stmt & MS & Pat & none & -- & D \\
Peng et al. \cite{p768} & 2025 & repo & RW & Ver & none & -- & D \\
Mono \cite{p951} & 2025 & stmt & MS & Pat & none & Pub & B \\
Yuan et al. \cite{p974} & 2025 & file & MS & Tool & Ded & Pub & D \\
LLM4CVE \cite{p988} & 2025 & stmt & MS & Pat & none & Pub & B \\
SPVR \cite{p195} & 2026 & func & MS & Pat & Ded & -- & D \\
Irsan et al. \cite{p220} & 2026 & stmt & RW & Ver & Prj & -- & B \\
VulnResolver \cite{p38} & 2026 & repo & RW & Ver & none & -- & D \\
Ananbeh et al. \cite{p669} & 2026 & func & RW & Ver & none & Pub & B \\
\midrule
\textbf{T4 Tests, PoCs \& exploits} &  &  &  &  &  &  &  \\
Ghera \cite{p1546} & 2017 & mod & SY & Ver & none & Pub & D \\
Magma \cite{pseedmagmagroundtruth} & 2020 & repo & IG & Ver & n/a & Pub & D \\
Afrose et al. \cite{p1449} & 2023 & func & SY & Ver & none & Pub & B \\
Zhang et al. \cite{p1458} & 2023 & func & RW & Ver & none & Pub & B \\
RedCode \cite{p1259} & 2024 & file & RW & Ver & none & Pub & D \\
Magneto \cite{p1277} & 2024 & mod & RW & Ver & Chr & -- & D \\
AgentDojo \cite{pseedagentdojodynamicenvironment} & 2024 & -- & SY & Ver & none & Pub & D \\
Ruan et al. \cite{pseedtoolemu} & 2024 & -- & IG & Ver & none & No & D \\
Zhao et al. \cite{p1001} & 2025 & repo & RW & Ver & none & -- & D \\
CyberGym \cite{p1109} & 2025 & repo & RW & Ver & Chr & -- & D \\
Bui et al. \cite{p1201} & 2025 & -- & -- & -- & none & -- & S \\
PBFuzz \cite{p2109} & 2025 & repo & RW & Ver & Con & Pub & D \\
VulnRepairEval \cite{p778} & 2025 & repo & RW & Ver & none & No & D \\
SecRepoBench \cite{p863} & 2025 & repo & RW & Pat & Con & -- & D \\
Dang et al. \cite{p898} & 2025 & -- & RW & Ver & Ded & -- & B \\
Qu et al. \cite{p113} & 2026 & file & IG & Ver & Ded & -- & D \\
SecureVibeBench \cite{p163} & 2026 & repo & RW & Ver & Con & -- & D \\
Chen et al. \cite{p170} & 2026 & func & RW & Ver & Con & Pub & D \\
Lee et al. \cite{p203} & 2026 & repo & RW & Ver & none & -- & D \\
PoCEvolve \cite{p2110} & 2026 & repo & RW & Ver & Ded & Pub & B \\
ChainDelta \cite{p2111} & 2026 & repo & RW & Ver & none & Pub & D \\
SABER \cite{bhu2026saber} & 2026 & repo & IG & Ver & none & Pub & D \\
Pu et al. \cite{p284} & 2026 & repo & RW & Ver & Chr & No & B \\
HardSecBench \cite{p454} & 2026 & mod & SY & Ver & none & Pub & D \\
FirmAgent \cite{p505} & 2026 & repo & RW & Ver & none & -- & D \\
\midrule
\textbf{T6 Traces \& logs} &  &  &  &  &  &  &  \\
Kozak et al. \cite{p766} & 2025 & -- & RW & Ver & none & -- & B \\
Kasu et al. \cite{p14} & 2026 & -- & SY & Tool & n/a & Req & D \\
SWE-chat \cite{bbaumann2026chat} & 2026 & -- & RW & Tool & n/a & Pub & B \\
\end{longtable}
%%% END-AUTOTABLES

\section{Discussion and Future Directions}\label{sec:discussion}

Three results structure the discussion that follows. \emph{First, realism and label evidence are associated with demonstrative strength, most clearly at its endpoints---though neither monotonically across every row, nor in company with availability, which runs the other way}
(Figure~\ref{fig:matrix}): tests, PoCs, and exploits (T4) are the only well-populated type for which majorities meet both criteria---because a run checks the triggered behaviour directly, and because these datasets more often use authentic targets with their context retained---while code-sample datasets (T1)---the largest category in both the auto-tagged corpus and the anchor set---are the weakest on realism, with one
of the 41 we deep-code using authentically real-world data---and that one only because its samples
are whole smart contracts rather than sliced functions. \emph{Second, two omissions recur across artifact types}: 49 of the 90 datasets where it applies never address train/test leakage
(Section~\ref{leakage-and-contamination}), and more than a quarter make no availability statement
at all (Section~\ref{availability}). \emph{Third, the explanatory and observational categories remain sparsely represented}: traces (T6) are represented by three deep-coded datasets, while reasoning (T5) has grown to eight since 2024, and no quality attribute has been the primary object of more than fourteen studies (Table~\ref{tab:t64b_studies_vs_datasets}). These observations are consistent with a common incentive: data that is inexpensive to label is easier to collect at scale than data that provides stronger evidence. A function labeled from a fix commit can be obtained from a diff, whereas an executable proof requires security expertise and a working environment. Accordingly, code-sample datasets are much more numerous than reusable executable datasets, while executable datasets are produced at smaller scale and have lower availability. The rest of this section uses these findings to develop recommendations for increasing the evidential support provided by vulnerability data.

\subsection{Good practice, attribute by attribute}\label{sec:good-practice}

We begin with what the corpus already shows can be done well. This subsection is addressed to the reader building or releasing a dataset, and answers a narrow question: for each quality attribute, what does good practice look like, and what should be done when the ideal is unaffordable? Section~\ref{sec:limitations} then states what remains out of reach even when this advice is followed, and Section~\ref{sec:future} sets an agenda for closing the difference. We give three tiers
throughout---a target, a fallback, and what to report either way---because most of the corpus
cannot reach the target, and guidance that only describes the target would be advice to do nothing.
Table~\ref{tab:datasheet} collects the reporting obligations into the datasheet
Section~\ref{sec:future} argues for.

\textbf{Realism.} \emph{Target:} draw samples from the situation in which the vulnerability actually
arose, with its context intact. SecureVibeBench states the principle most sharply---a fair
evaluation places the agent in the exact scenario where a human developer originally introduced the
flaw~\cite{p163}---and SecCodeBench-V2 shows it is achievable at function granularity provided the
function stays inside its project scaffold~\cite{p170}. \emph{Fallback:} where full context is
unaffordable, mine at the coarsest granularity the labeling budget allows rather than defaulting to
the function, and say what was discarded. \emph{Report either way:} provenance (human, generated, or
mixed), the granularity at which samples were cut, and whether any sample was validated by execution.
The realism value a reader can compute from those three is more useful than an author's assertion that
a corpus is real-world, which is how several records in this study were initially miscoded.

\textbf{Label evidence.} \emph{Target:} a label that something other than the labeler can check---an exploit that runs, an instrumented crash, a test that fails before the patch and passes after.
Magma achieves this by instrumenting known bugs so that triggering is directly
observable~\cite{pseedmagmagroundtruth}, and PBFuzz names the property that makes it work: proofs of
vulnerability are expensive to produce and cheap to verify~\cite{p2109}. \emph{Fallback:} where no
oracle exists, prefer two independent weak signals to one---SecCodeBench-V2 double-reviews every
case with security experts~\cite{p170}, and where tools must be used, cross-checking two analyzers
catches more than trusting one~\cite{p857}. \emph{Report either way:} the oracle, not the outcome.
``Verified'' is not a label provenance; ``confirmed by re-running the PoC'' is. This single change would let a consumer separate the 51 verified datasets in our corpus into
those confirmed by execution and those confirmed by manual review.

\textbf{Scale.} \emph{Target:} there is no universal target. The relevant goal is the largest corpus the chosen oracle can support without degrading quality. Scale appears in these cases to be constrained by the label oracle (Section~\ref{scale}). \emph{Fallback:} not applicable---scale needs no fallback, it needs a reported scaling curve. Stop adding volume once marginal utility plateaus or quality degrades; one augmentation study observed such a plateau at roughly 5k synthetic samples in its own setting~\cite{pseedvulscriberexploringrag}. \emph{Report either way:} the
count with its unit, and the funnel that produced it. VulnRepairEval is the model here: reporting that
2{,}500 sources and 448 CVEs yielded 23 usable instances says far more about the state of executable
evidence than the number 23 alone~\cite{p778}.

\textbf{Diversity.} \emph{Target:} coverage reported on all four facets---CWE, project, language,
and class balance---and chosen rather than inherited. JavaVulBench spans 225 CWE categories over
700+ projects~\cite{p412} and SecCodeBench-V2 covers five languages from industrial
code~\cite{p170}; both are deliberate choices, not the residue of what a mining query returned.
\emph{Fallback:} a single-language or single-CWE-family dataset is entirely legitimate, most of the corpus is one, provided the scope is stated as a scope rather than presented as general. The 45 single-language datasets in our corpus are not a problem individually; the problem is that few of
them say so. \emph{Report either way:} class balance against the deployment base rate. PrimeVul
reports roughly one vulnerable function in 33 and argues that accuracy and F1 misrepresent
performance whenever a dataset departs from it~\cite{pseedprimevul}; a balanced dataset that does not
disclose its balance invites exactly that error.

\textbf{Leakage and contamination.} \emph{Target:} the strongest split the data supports, plus an
explicit contamination check when a pretrained model is the consumer. Among the datasets we deep-code, JavaVulBench is the only one that ships this combination: five split strategies---random, project-disjoint,
temporal, deduplicated, unseen-CWE-family---shipped together with a per-model contamination
audit~\cite{p412}. \emph{Fallback:} a chronological split adds no per-sample labeling effort and is a high-value default for temporal generalization; it mitigates pretraining contamination only where the evaluation portion postdates the consumer model's training cutoff. Where no split
applies at all, guard by construction, as the five contamination-controlled datasets in our corpus
do with post-cutoff, proprietary, mutated, or cutoff-argued data~\cite{p163,p168,p170,p2109,p863}. \emph{Report either
way:} the split strategy, or a statement that none applies and why. This is the cheapest obligation
in the table and the most frequently skipped: 49 of 90 applicable datasets report nothing.

\textbf{Availability.} \emph{Target:} a resolvable, versioned link at submission---not on
acceptance, and not on request. \emph{Fallback:} where licensing or disclosure genuinely prevents
release, publish the construction pipeline instead, so the artifact can be rebuilt even if it cannot
be shared. \emph{Report either way:} the status, explicitly. No availability statement is far more common than an explicit refusal---25 of 93 datasets make no availability statement at all, against 5 that state one---and non-reporting is the worse of the two, because it cannot be planned around. Executable artifacts deserve particular emphasis: they share the lowest public-availability rate with patch datasets, at 12 of 24 against 7 of 14, and are the costliest to rebuild when withheld.

\begin{table}[t]
\caption{A datasheet for vulnerability data: the minimum a release should state, per attribute, and
what a consumer cannot assess without it. Most can be reported without new research where the information already exists; producing missing validation, contamination, or provenance evidence may require additional engineering.}
\label{tab:datasheet}
\small
\begin{tabularx}{\linewidth}{@{}p{2.2cm}XX@{}}
\toprule
Attribute & State at minimum & Otherwise a consumer cannot assess \\
\midrule
Realism & provenance (human/generated/mixed); granularity of the sample; whether any sample was
execution-validated & context fidelity, or how far results are likely to transfer beyond sliced code \\
Label evidence & the oracle that produced the label, not the confidence in it & whether an independently checked label was confirmed by execution or by manual review, or was only tool-derived \\
Scale & count with unit, and the funnel that produced it & how much was discarded, and therefore how representative the remainder may be \\
Diversity & CWE, project, language, and vulnerable:benign balance; scope stated as scope & a score relative to language, CWE and project scope, and to class prevalence \\
Leakage & split strategy, or that none applies and why; contamination check for pretrained consumers
& train/test independence, or what contamination controls were applied \\
Availability & status and a resolvable link, or the pipeline if the artifact cannot be shared & whether the artifact can be obtained for reuse or reproduction \\
\bottomrule
\end{tabularx}
\end{table}

Two things are worth noticing about Table~\ref{tab:datasheet}. Several of the six obligations require little beyond writing down what the builder already knows at release. Others---supplying missing validation, contamination, or provenance evidence---can require substantial engineering, and we do not claim otherwise. And the consequences in the third column are not
hypothetical---each is a failure this study documents, from performance that does not transfer
(Section~\ref{realism}) to accuracy that reflects base rate rather than capability
(Section~\ref{diversity}) to results that cannot be rebuilt (Section~\ref{lim:reproducibility}).
That is the practical form of the argument in Section~\ref{attributes-not-independent}: convention is an important constraint on data quality alongside cost.

\subsection{Limitations of Existing Work}\label{sec:limitations}

Section~\ref{survey-result-data-quality-rq2-the-core} describes what vulnerability data \emph{is like}, and the practices above describe how to build it better. This subsection states what the field \emph{cannot do} regardless, on the data foundation it currently has. We name seven limitations. They are not a list of individual
papers' shortcomings---most of the work cited below is the work that identified the limitation---but
properties of the field's current data foundation that no single contribution can remove.

\subsubsection{Most vulnerability labels lack an affordable independent oracle}\label{lim:oracle}
The field lacks a general-purpose, affordable oracle that establishes vulnerability across the relevant categories. Every route in use trades
soundness against cost. Devign's authors state the founding version of the problem: previously
released vulnerable-function datasets were labeled by static analyzers and are simply not accurate,
while manual labeling demands security expertise and therefore cannot
scale~\cite{pseeddevigneffectivevulnerability}. Two decades of practice have not dissolved that
dilemma; they have distributed it. Patch-derived labels inherit the tangled-commit error that
CVEfixes names from inside the practice~\cite{pseedcvefixesautomatedcollection}; scanner-derived
labels inherit the scanner's false positives~\cite{p826,p857}; and Mono shows the residue is not even noise but \emph{undecidability}: in the patch-derived benchmark it evaluates, for roughly one CVE in six, no procedure restricted to the provided static-code context could reliably recover the intended label~\cite{p951}.

The consequence extends past dataset construction into evaluation. For generated code there is no
accepted security oracle the way pass@$k$ exists for functional correctness~\cite{p1520}, so the field
falls back on scanner-plus-inspection over hand-built scenarios, neither automatable nor comparable
across papers. Where a narrowly defined property admits a decisive oracle, label ambiguity largely disappears and very large-scale measurement becomes possible: package
hallucination admits exact counts at 576,000-sample scale because a package either exists in a
registry or does not~\cite{p1222}. For general software vulnerabilities the limitation combines cost with semantic and coverage constraints, which is why it has proved so durable.

\subsubsection{The unit of data discards the evidence}\label{lim:context}
To make automatic labeling tractable, the field slices code to the function: 50 of the 111 deep-coded
papers work at that granularity. The slice is what makes labeling affordable, and it is also what
removes the calling context, build configuration, and inter-procedural flow in which a vulnerability
actually manifests. Function-sliced datasets therefore cannot represent vulnerabilities whose manifestation depends on the interprocedural or environmental context that slicing removes.

The evidence that this is a representational limit rather than a difficulty is that models exploit
the residue. ReVeal finds detectors keying on dataset artifacts such as specific variable and
function names rather than on the cause of the vulnerability~\cite{pseedreveal}, and a later study
attributes the same collapse to samples that are ``context-independent with known
patterns''~\cite{p1276}. When context is restored, the missing population becomes visible: Mono
recovers 89\% of inter-procedural vulnerabilities that function-sliced pipelines had
dropped~\cite{p951}, and 21 of APPatch's 97 zero-day samples are interprocedural, with manifestation
and patch locations in different functions~\cite{p1403}. A dataset built one function at a time
cannot contain these cases, however large it grows.

\subsubsection{Seed-derived generators inherit their source's coverage}\label{lim:generation}
Synthesis is the standing answer to scarcity, and it is bounded in a way that more compute does not relieve: a seed-derived generator whose edit vocabulary is learned only from its source corpus is limited to the patterns represented there, unless additional knowledge or generation mechanisms expand the space. VGX states this directly---learned edit patterns cannot exceed the small vulnerability dataset
they were mined from, so class coverage is inherited rather than
expanded~\cite{pseedvgxlargescale}---and VulGen's confinement to single-statement injections is the
same limit in structural form~\cite{pseedvulgenrealisticvulnerable}. Over 65\% of vulnerabilities in
the largest real-world corpora require multi-line changes that such generators cannot
express~\cite{pseedexploringimprovingreal}.

Volume does not compensate. VinJ reaches 686k samples at a 69\% correctness rate, none validated by
execution, so roughly a third of the corpus is mislabeled by construction~\cite{p1230}; in one augmentation setting, gains plateaued past roughly 5k generated samples, showing that additional volume need not carry additional signal~\cite{pseedvulscriberexploringrag}. The deeper problem is evidentiary rather than statistical: the generators in our corpus typically argue their realism \emph{indirectly}, by showing that detectors trained on
their output improve, rather than demonstrating that any generated sample is exploitable. Adversarial
augmentation makes the circularity explicit---perturbing embeddings improves robustness without
adding any real evidence about the rare classes it is meant to cover~\cite{p429}.

\subsubsection{Reported performance does not measure capability}\label{lim:evaluation}
A majority of the applicable datasets we coded report no effective leakage guard (Section~\ref{leakage-and-contamination}), and for LLM consumers it is compounded by pretraining contamination, which is difficult to measure directly. Because training corpora are typically undisclosed, contamination often has to be assessed indirectly; perturbation sensitivity is one such diagnostic, and it varies enormously across benchmarks and therefore makes
cross-benchmark claims about LLM security capability non-comparable~\cite{p378}. The signature is
visible where it can be checked: Copilot reproduces original vulnerable code roughly a third of the
time and more often for \emph{older} vulnerabilities~\cite{p1494}, and models scoring well on public
benchmarks collapse on post-cutoff zero-day samples~\cite{pseedassessingimprovingprompting}.

The metrics compound the problem. Accuracy and F1 misreport performance whenever a research dataset
departs from the real base rate of vulnerable code, as PrimeVul argues~\cite{pseedprimevul}; one model
swings by over 36\% F1 between balanced and imbalanced versions of the same real-world
data~\cite{p1276}; and in repair, textual similarity is not a security oracle---patches compile and
score well on BLEU while failing to fix the vulnerability or introducing new insecure
patterns~\cite{p669}. Two studies show the resulting fragility directly: detector rankings invert
between benchmark suites, with artificially generated vulnerabilities proving easier to detect than
real ones~\cite{pseedevaluatingcomparingmemory}, and when secure-code generation is evaluated jointly
for security and functionality with multiple detectors rather than one, many published techniques \emph{degrade} the base model by more than 50\%, showing that reported gains can depend strongly on the evaluation protocol~\cite{p888}.

\subsubsection{Coverage is siloed, and the silos are unrepresentative}\label{lim:coverage}
Quality-labeled vulnerability data is fragmented by language and weighted toward C/C++, a
concentration our coding quantifies: C/C++ appears in 27 of 93 datasets, more than any other language, and accounts for 21 of the 45 single-language datasets; only 7 datasets span three
languages or more
(Section~\ref{diversity}). This is not a gap that broader collection would close, because the
selection criteria themselves are narrowing. Big-Vul admits only Git-hosted, CVE-linked projects and
drops entries that cannot be mapped to a CVE~\cite{pseedbigvul}; patch-identification models trained
on NVD-linked data lose up to 90\% F1 in the wild because publicly disclosed patches are a biased
sample of real ones~\cite{p220}. Widely used mined datasets therefore sample publicly disclosed and well-linked vulnerabilities rather than vulnerabilities in general. Within each silo, coverage is long-tailed. PatchDB's
NVD-derived component is explicitly imbalanced enough that its authors synthesize control-flow
variants to populate it~\cite{pseedpatchdblargescale}; repair accuracy is CWE-shaped, tracking corpus
coverage rather than difficulty~\cite{pseedvulrepairt5based}; and scope is sometimes capped by
design at the CWE Top-25~\cite{p195}. Outside C/C++, evaluation infrastructure is thin even where
deployment is heavy---Java's prior standard offered 79 vulnerabilities~\cite{p669}---and ecosystems
such as smart contracts, blockchain clients, and IoT firmware are served by isolated one-off
efforts~\cite{p974,p2111,p505}, several reporting no dataset size, provenance, or labeling protocol
at all~\cite{p287}.

\subsubsection{Reproducibility remains limited}\label{lim:reproducibility}
The single study that measured reproducibility end-to-end found a funnel rather than a rate: of the
deep-learning vulnerability detectors reviewed, only 25.5\% released a tool; of those reproducible,
14.3\% were partially and 28.6\% fully replicable on a different real-world dataset, and 57.1\% could not process the
samples, or could process only part of them~\cite{p1433}. Tool availability and cross-dataset replicability therefore limit how much reported capability can be independently checked. We hold this study
to the same standard we ask of the field: Appendix~\ref{app:protocol} states the
queries, retrieval parameters, and screening dispositions behind our own corpus, and
the coded records and analysis scripts are released with it.

Two structural factors sustain this. First, availability is not treated as a reportable property: 25 of 93 datasets make no statement about it, and executable artifacts share the lowest public-availability rate among the major types, at 12 of 24, and are the costliest to reconstruct when withheld
(Section~\ref{availability}). Second, even available artifacts are not comparable, because there is no
shared construction protocol. Thirteen malicious-code prompt corpora were each built under different
inclusion criteria, licensing, and validation standards, with per-prompt reliability typically
unreported, so consolidating across them requires reconstructing each corpus's criteria from
scratch~\cite{p232}; a review of 114 primary studies finds the same fragmentation leaves practitioners
with no end-to-end quality-assurance guidance~\cite{p636}. Exploit-generation research compounds
both, since few studies release usable tools~\cite{p1201}.

\subsubsection{Three failure modes the corpus cannot support work on}\label{lim:blindspots}
Some limitations are not degrees of quality but absences of scale and grounding. For each of the
three failure modes below the corpus contains \emph{something}---three trace datasets, eight reasoning datasets, a provenance field on every record---and in each case the scale or the grounding is short
of what the failure mode demands.

\textbf{What an agent does.} Agents emit insecure actions in 21\% of trajectories during ordinary
operation, with no attacker involved~\cite{p766}. This lives in the action stream---which file was
read, which endpoint called, which artifact fetched unverified---not in any diff, so no code-centric
dataset can surface it at any scale. Building the trajectory-level detector reported in~\cite{p766} required hand-labeling 500 trajectory steps, because no labeled corpus of agent behavior existed.

\textbf{Why code is vulnerable.} Reasoning datasets have begun to appear rapidly
(Section~\ref{res:t5}), but the corpora built at scale are model-generated annotations checked by
another model, and the one expert-authored set stops at 174 cases. Consumers can therefore check
\emph{that} a label was assigned, and increasingly read a machine's account of \emph{why}, without
any independent ground truth for that account---which is precisely the check that would catch the
undecidable cases of Section~\ref{lim:oracle}.

\textbf{Where code came from.} We code provenance as human, AI-generated, or mixed for every
record, but no dataset carries that distinction in a form a consumer could act on---no attestation
travels with the artifact, and nothing in the distribution channels these artifacts travel through
carries integrity evidence either. A poisoned code model is distributable as an ordinary open-source artifact,
survives state-of-the-art defenses, and produces vulnerable code 95\% of the time under a stylistic
trigger while losing under 5\% on functional benchmarks~\cite{p274}. A single poisoned sample in a
retrieval knowledge base drives 48\% of generated code to contain vulnerabilities~\cite{p1035}. A
public marketplace of 631{,}813 agent skills ships with no mandatory security
review~\cite{p113}. Consumers cannot distinguish generated from mined code, because no released artifact we coded carries sample-level authorship provenance with the data. Separately, the absence of integrity attestation leaves no way to verify that a distributed artifact reached its consumer untampered---two gaps we can evidence across our corpus, though not, on this evidence, beyond it.

\subsection{Future Directions}\label{sec:future}

Each direction below names the limitation from Section~\ref{sec:limitations} it addresses, what
already exists that points the way, and the specific problem left open. They are ordered from the
most tractable to the most speculative, and we return to that ordering at the end.

\subsubsection{Make executable validation a construction step}
\emph{Addresses Section~\ref{lim:oracle} (no general-purpose affordable oracle) and
Section~\ref{lim:reproducibility}.} A runnable trigger carries a direct check of reachability in the tested environment,
and the corpus measures precisely how scarce it is: 78.9\% of CVEs have no publicly available proof-of-concept~\cite{p898}, 70.2\% of CVE patches ship no test files~\cite{p2110}, and building one
exploit-based benchmark required processing roughly 2{,}500 sources and 448 CVEs to retain
23 usable instances~\cite{p778}.

What has changed is that producing such evidence is no longer purely manual. Instrumented ground
truth re-introduces real bugs into real targets so that triggering is directly
observable~\cite{pseedmagmagroundtruth}; agent-driven reproduction turns a patch into a working
exploit for over half of a set of Linux-kernel N-days~\cite{p284}, for 57 of 129 Magma
CVEs~\cite{p2109}, for 16 of 25 Ethereum vulnerabilities~\cite{p2111}, and, without a vulnerability
report at all, from the patch alone~\cite{p2110}. And as Section~\ref{res:t4} notes, the one
executable benchmark that reaches four figures does so because PoC production already sat upstream of
it~\cite{p1109}, which is the arrangement this direction proposes to generalize.

The open problem concerns both the remaining reproduction capability and the integration of that capability into dataset construction. Every result above is a
standalone contribution; none is a step in someone else's dataset pipeline. Concretely, we see three
tasks. First, report a \emph{validation yield} for every dataset that claims real vulnerabilities, the fraction of samples for which a triggering input exists, so the currently invisible difference
between a mined corpus and a validated one becomes a number. Second, run patch-to-PoC generation at
collection time rather than as a downstream evaluation, converting the 70.2\% of patches that ship no
tests into candidates. Third, publish the failures: the CVEs that resisted automated reproduction are
the sharpest available characterization of what remains hard, and they are currently discarded.

\subsubsection{Validate generated data by execution where it applies}
\emph{Addresses Section~\ref{lim:generation}.} The generators we deep-code primarily argue their realism \emph{indirectly}, by showing that detectors trained on their output improve. That argument is
circular when the detector is evaluated on data from the same distribution, and it hides defects a
direct check would catch: an industrialized injector reaches 686k samples at a 69\% correctness
rate~\cite{p1230}, and augmentation gains plateau past roughly 5k generated
samples~\cite{pseedvulscriberexploringrag}, which is what one would expect if additional samples
carry no new information.

The fix follows from the previous direction: a generated sample should ship with a triggering input,
or be labeled as unvalidated. Two further constraints deserve explicit measurement. Generators
remain largely confined to single-statement injection while over 65\% of real vulnerabilities require
multi-line changes~\cite{pseedexploringimprovingreal}, so \emph{edit span} is a reportable property
of a generator, not an implementation detail. And the gap between synthetic and real performance---79.40\% against 10.03\% for the same neural editors~\cite{pseedgeneratingrealisticvulnerabiliti}---should be reported by construction, as a train-synthetic/test-real number, rather than discovered
later by someone else.

\subsubsection{Preserve context: move mining up the granularity ladder}
\emph{Addresses Section~\ref{lim:context}.} The function slice is what makes automatic labeling
affordable and what removes the calling context, build configuration, and inter-procedural flow in
which a vulnerability manifests. The cost is measurable: restoring context recovers 89\% of
inter-procedural vulnerabilities that function-sliced pipelines drop~\cite{p951}, and detectors
trained on sliced corpora key on dataset artifacts rather than on the cause of the
vulnerability~\cite{pseedreveal}.

Repository-level benchmarks show the target is reachable---318 completion tasks across 27
repositories~\cite{p863}, 105 tasks whose repositories average 2{,}845 files~\cite{p163}, 183
validated instances in two JavaScript engines~\cite{p203}---but all are \emph{evaluation}
artifacts, built at hundreds of instances. The open problem is repository-level data at
\emph{training} scale: mining that retains build configuration and cross-function flow, and a label
that survives the transition. Until that exists, these anchors train on slices and evaluate on repositories, which is the mismatch Section~\ref{sec:conclusion} identifies as the
current trajectory.

\subsubsection{A security oracle for generated code}
\emph{Addresses Section~\ref{lim:evaluation}.} As LLMs and agents come to dominate the consumption recorded in our anchors (Section~\ref{survey-result-consumption-and-evaluation-rq3}), they also become
prolific \emph{producers} of code whose security must be judged, and here the field lacks even a
metric. There is no accepted oracle for the security of generated code the way pass@$k$ exists for
functional correctness~\cite{p1520}, and the two properties come apart sharply: 61\% of one agent's
solutions are functionally correct and only 10.5\% secure~\cite{p1001}, while 4.3--6.0\% of
functionally correct agent patches are vulnerable with no attacker present~\cite{p768}.

Benchmarks that judge security and functionality \emph{jointly} on the same executable artifact are
the right shape~\cite{p863,p170}, and the honest statement of their limit comes from one of them:
passing the tests does not imply the absence of vulnerabilities, and a whole class of issues---weak
cryptography, hardcoded credentials, information leakage---cannot be adjudicated by tests at
all~\cite{p170}. A usable oracle therefore needs two tiers: executable checks where they apply, and a
declared, auditable fallback where they do not. Reporting which tier decided each verdict would make secure-code-generation results substantially more comparable; today a single scanner's judgment and a
running exploit are reported as the same kind of evidence, and evaluating jointly with multiple
detectors turns many published gains into regressions~\cite{p888}.

\subsubsection{Provenance as a first-class field}
\emph{Addresses Section~\ref{lim:blindspots}.} The attacks catalogued there---a poisoned model
that survives current defenses, a single poisoned retrieval sample, an unreviewed skill marketplace---share one precondition: no released artifact we coded carries sample-level, machine-readable provenance, so no consumer can tell mined code from generated. Authorship provenance is not integrity evidence---an authenticated human-written sample can still be malicious---and attestation is a separate requirement we return to below.

Our taxonomy records provenance---human, AI-generated, or mixed---and we could assign it for every
record from the paper's own description. What no dataset carries is that distinction \emph{at sample
level}, in a machine-readable form that travels with the artifact, so a consumer holding the data
cannot tell which samples were model-written. Adding the field is inexpensive where the information is already known, though preserving it at sample scale may require additional infrastructure, and integrity attestation---evidence that an artifact reached the consumer untampered---is a distinct problem that provenance alone does not address. Closing the provenance gap nonetheless has compounding value: a
consumer who can separate mined from generated code can measure whether training on AI-written code
degrades security, which is currently unanswerable at corpus scale even though a quarter to a third
of AI-generated snippets merged into open-source projects carry weaknesses~\cite{p850}.

\subsubsection{Agent behavior as data}
\emph{Addresses Section~\ref{lim:blindspots}.} The least-served artifact is the record of what an
agent \emph{does}, not what it writes, and the failure rate reported there is high enough that the absence of such records is a major limitation for training and evaluating trajectory-level safety detectors.

Executable agent environments are the beginning of this
data~\cite{p1259,pseedagentdojodynamicenvironment,pseedtoolemu}, and two recent contributions release
records rather than only environments: a corpus of coding-agent interactions with real
users~\cite{bbaumann2026chat} and a benchmark of operational agent safety in stateful
settings~\cite{bhu2026saber}. With the hand-labeled trajectory corpus~\cite{p766}, the primary T6 row contains three datasets, and three further datasets record traces as secondary artifacts. Vetting autonomous agents \emph{before} they act on real systems at scale would be enabled by a trajectory-level, safety-labeled corpus large enough that a detector can be trained rather than prompted---and labeling cost is a major obstacle: the trajectory-level detector reported in~\cite{p766} required hand-labeling 500 steps to build.

\subsubsection{A reporting standard for vulnerability data}
\emph{Addresses Section~\ref{lim:evaluation} and Section~\ref{lim:reproducibility}.} The small number of studies represented in the studies column (Table~\ref{tab:t64b_studies_vs_datasets}) reflects a reporting gap as well as a research gap. Leakage and availability go unstated often enough that basic comparability is lost, thirteen
malicious-prompt corpora were each built under a different and incompatible
protocol~\cite{p232}, and only a quarter of studied detection tools were released at
all~\cite{p1433}.

A lightweight ``datasheet'' for vulnerability data would make four things mandatory rather than
optional: split strategy or contamination guard, label provenance, code provenance, and availability.
Section~\ref{sec:good-practice} works that proposal out attribute by attribute, and
Table~\ref{tab:datasheet} states it as a checklist.
JavaVulBench is a workable template---five split strategies plus a per-model contamination
audit~\cite{p412}---and the marginal labeling cost is low, which is what makes this the most tractable item here. As Section~\ref{attributes-not-independent} argues, leakage is the one attribute that
depends on no oracle and asks for a partitioning decision rather than new evidence;
that a majority of the applicable datasets we coded still skip it suggests reporting convention is an important constraint alongside resource cost.

\subsubsection*{Relative tractability} These directions differ sharply in what they require. Where the underlying information already exists, the reporting standard needs no new technique or data and would make the other attributes substantially more assessable. Validation yield and edit span are likewise inexpensive to report for pipelines that already measure them, and recording provenance is inexpensive where it is known. Repository-scale
training data and trajectory-level safety corpora are the expensive items, each requiring infrastructure that does not yet exist, and each demanding substantial engineering and labeling effort. That ordering suggests several immediate data-quality gaps can be narrowed through reporting practice, while others still require new research and infrastructure.

\section{Threats to Validity}\label{threats-to-validity}

The findings depend on human judgments applied to descriptions in the source papers.
We state the threats in the standard four categories; Appendix~\ref{app:validity} carries the
supporting counts, the triage record, tagger accuracy by period, and the screening audits.

\subsection{Construct validity}\label{construct-validity}
The six quality attributes and their values are ours, not the papers'
(Section~\ref{sec:tax}), so every graded value is an interpretation. We mitigate this with
item-level definitions, a worked example per attribute, and the rule that no affirmative
graded value is recorded without a verbatim span from the source, 729 spans across 111
papers. That yields auditability rather than reliability: a reader who disputes a judgment
can see exactly what it rested on, but we have not established that an independent coder
would reach the same one. Absence values are the exception, since they cannot be supported by a quotation; we checked them by targeted full-text scans, and they measure reporting practice as
much as underlying quality. We also assign one primary artifact type per paper, so datasets
that bundle several are under-counted in rows other than their own.

\subsection{Internal validity}
Records were drafted by an LLM from full text and then verified against the papers by the
authors: 80 checked field by field, and 31 from a second pipeline screened against four
conflict signals, adjudicated by three authors, and then checked field by field as well.
Verification found one systematic fault, a pipeline default on \emph{provenance}, since
corrected corpus-wide, and one record whose values had been drawn from a sentence describing
a different dataset. Where three authors judged the same flags, 94 of 99 verdicts upheld the
coded value. Because that measures verification rather than coding from a blank form, we
also report how much disagreement the conclusions absorb: the claim that executable artifacts are the only well-populated type meeting both criteria by a majority survives two adverse record recodes, and the leakage finding survives eight.

\subsection{External validity}
The deep-coded sample is purposive, not random: 111 of 1{,}522 included papers, selected as
foundational datasets, high-impact recent contributions across all six artifact types, and
every paper reporting a finding about data quality. It is the right sample for characterizing
the field's \emph{anchor} artifacts and the wrong one for estimating a population mean, and the direction of the resulting bias is not identifiable: anchor papers are generally better documented than the median paper, while the deliberate inclusion of every paper reporting a data-quality finding also enriches the set for documented failures.
Our original screen also filtered within tagger-assigned types, so a mistyped paper could
never be reconsidered; we re-screened all 808 excluded records, which returned 217, and a
second-reader sample of the remainder put screening recall within the retrieved candidate
pool at 98.6\%. That bounds false exclusions among retrieved candidates and says nothing about retrieval recall, which we do not measure: several APIs imposed per-query caps, one was intermittently unavailable, and the query vocabulary necessarily favours papers whose titles and abstracts use the terms we searched. Named-artifact seeding, citation chasing, supplemental queries, and a recall-oriented search by two of the authors mitigate this without eliminating it, and we searched in English only. The tagger that assigns corpus-wide
types agrees with the manual type on 70.7\% of the papers carrying one and its recall is not
stationary, so no compositional claim in this study rests on it.

\subsection{Conclusion validity}
One row of the matrix rests on three deep-coded datasets, and we treat its emptiness rather
than its values as the finding; the reasoning row was in the same position until a corpus-wide re-screen and a subsequent pass for data-quality studies took it from one dataset to eight. Percentages are computed over 93
contributed datasets, so a single reclassification moves a figure by about a point, and the
per-band counts in Section~\ref{sec:trajectory} are smaller still. Where a claim depends on a
handful of records we give the fraction, not only the percentage.

\section{Conclusion}\label{sec:conclusion}

We have studied software vulnerability analysis from the position of its data, deep-coding
111 anchor papers from a systematically assembled corpus of 1{,}522. The organizing result is an
\emph{evidence ladder}: artifacts differ in whether they \emph{assert} that a vulnerability exists or \emph{demonstrate} that it does, and the coded quality profile varies with that difference, most clearly at its ends---though part of that association follows from the operational definitions of executability and realism rather than from independent measurement. Among the types with enough datasets to carry a rate, executable
artifacts are the only one for which majorities satisfy both realism and label evidence, at
16 of 24 and 23 of 24, while code samples, the largest category in the corpus and in the
anchor set, manage one of 41 on realism. Three limitations remain prominent among the datasets we deep-code: most applicable datasets report no effective leakage guard, a quarter make no availability statement, and artifacts for explaining verdicts or recording agent behaviour remain scarce. Several of these problems have already been identified in prior work, but our synthesis shows that they remain common in practice. The gap between documented best practices and observed practice is especially clear for low-cost measures such as chronological splitting, which requires a partitioning decision rather than additional labeling effort. This gap becomes more important as consumers become autonomous and act on data rather than only presenting results to developers. It motivates a shift from data that asserts vulnerabilities to data that demonstrates them.

\clearpage
\appendix
\begingroup
\centering
\vspace*{0.5\baselineskip}
{\Large\bfseries Appendices\par}
\vspace{0.5\baselineskip}
\endgroup
%\noindent The appendices record operational detail supporting Sections~\ref{survey-methodology} and~\ref{threats-to-validity}. No result in the body depends on reading them.

\section{Search and Screening Protocol}\label{app:protocol}

This appendix records the operational detail behind
Section~\ref{survey-methodology}: the queries as issued, the retrieval parameters,
the de-duplication rule, and how the screening dispositions partition the candidate
pool. The three harvesting phases ran between 8 and 17 July 2026, and the corpus was
frozen for coding on 28 July 2026; the two trace-data papers added in
Section~\ref{res:t6} were coded on 3 August 2026 after a reviewer noted they had been
cited but not coded. The harvesting and screening
scripts are released with the corpus, so each figure below can be recomputed rather
than taken on trust.

\subsection{Phase 1 --- topic harvesting}\label{app:phase1}

Twenty-three phrase queries were issued across four tagged families. Each query was
sent to the arXiv API, the DBLP publication API, and, best effort, the Semantic
Scholar Graph API, which was disabled for the remainder of a run after repeated
HTTP~429 responses. Requests were spaced 3\,s apart for arXiv and 1\,s for DBLP.

\begin{itemize}
\item \textbf{Data (6).} \texttt{software vulnerability dataset}; \texttt{vulnerability
  data generation}; \texttt{vulnerability data augmentation}; \texttt{vulnerability
  benchmark}; \texttt{CWE classification}; \texttt{vulnerability label noise}.
\item \textbf{AI for security (8).} \texttt{vulnerability detection deep learning};
  \texttt{vulnerability detection large language model}; \texttt{vulnerability
  localization}; \texttt{vulnerability repair large language model}; \texttt{automated
  vulnerability patching}; \texttt{proof of vulnerability generation}; \texttt{exploit
  generation large language model}; \texttt{LLM fuzzing}.
\item \textbf{Security of AI-generated code (4).} \texttt{LLM generated code security};
  \texttt{secure code generation}; \texttt{GitHub Copilot security};
  \texttt{AI-generated code vulnerabilities}.
\item \textbf{Agent behaviour (5).} \texttt{LLM agent security}; \texttt{coding agent
  safety}; \texttt{package hallucination}; \texttt{prompt injection agent}; \texttt{LLM
  agent sandbox}.
\end{itemize}

\noindent Retrieval parameters were: arXiv, exact-phrase field search
(\texttt{all:"\textit{phrase}"}), at most 150 results per query, sorted by submission
date descending; DBLP, \texttt{format=json} with \texttt{h=300}; Semantic Scholar,
\texttt{limit=100} with a \texttt{year=2020--} filter. This phase applied a 2020 floor,
one year below the review window, so that the 2016--2019 range and the pre-2016
classics enter through Phase~2 rather than being retrieved twice.

\subsection{Phase 2 --- named-artifact seeding and backfill}\label{app:phase2}

Two query sets ran in this phase. Fourteen topic queries with a \textbf{2016} floor
filled the 2016--2020 gap through DBLP and arXiv: the six data queries of Phase~1 plus
\texttt{code vulnerability dataset}, \texttt{security patch dataset},
\texttt{automated vulnerability repair}, \texttt{learning based vulnerability
detection}, \texttt{memory error vulnerability detection}, \texttt{fuzzing benchmark},
\texttt{exploit generation}, and \texttt{vulnerability localization}.

Nineteen named-artifact queries ran with \emph{no} year floor, so that foundational
resources predating the window could enter: SARD, Juliet, VulDeePecker, SySeVR,
$\mu$VulDeePecker, Draper VDISC, Devign, Big-Vul, D2A, Russell et al., SATE, CVEfixes,
PatchDB, Magma, VulDeeLocator, ReVeal, code-property-graph work, and two generic
queries on dataset construction and C/C++ detection benchmarks. In total 37 artifacts
were curated by name; 31 entered the pool as new records and 6 had already been
retrieved by topical search. Only titles not already present were merged, leaving
existing rows untouched.

\subsection{Phase 3 --- supplemental retrieval and enrichment}\label{app:phase3}

Exact-phrase search returned few results for the newer areas, so a supplemental pass
added ten arXiv conjunctive queries---\texttt{copilot AND security}, \texttt{generated
code AND vulnerabilities}, \texttt{code generation AND security}, \texttt{coding agent
AND security}, \texttt{LLM agent AND safety}, \texttt{agent AND prompt injection},
\texttt{exploit AND large language model}, \texttt{security patch AND dataset},
\texttt{proof of concept AND vulnerability}, \texttt{chain-of-thought AND
vulnerability}---together with six DBLP retries for queries that had failed. Records
were then enriched with venue, year, citation count, and abstract through OpenAlex,
Unpaywall, and Semantic Scholar, and extended by backward citation chasing from the
seeds.

\subsection{De-duplication}\label{app:dedup}

Titles were reduced to a normalised key by lower-casing and collapsing every run of
non-alphanumeric characters to a single space; keys shorter than 15 characters were
discarded as too weak to match on. Records sharing a key were merged, with the first
source retaining attribution and empty fields filled from later duplicates. DOI
equality was treated as a match regardless of title. Across all three phases the
sources contributed 2{,}240 records and merging removed 128, leaving \textbf{2{,}112}
candidates.

\subsection{Screening dispositions}\label{app:screening}

Screening applied transparent keyword rules over title and abstract. The rules are
deliberately asymmetric: a record is excluded only when an out-of-scope domain signal
fires \emph{and} no in-scope signal does, so anything ambiguous is retained as
\texttt{maybe} for human review rather than dropped. Nine out-of-scope domain patterns
were used, each recording its own reason: network intrusion detection and traffic
analysis; malware and ransomware analysis; phishing and content abuse; adversarial
machine learning outside a code context; vision and other modalities; federated
learning and ML privacy; hardware and side-channel security; other AI applications;
and generic chatbot jailbreaking with no software context.

Every candidate received exactly one disposition, recorded in a
\texttt{decided\_by} field: 1{,}171 by automated rule, 668 by LLM-assisted abstract
screening where the rules left the record ambiguous, 236 by independent screening
from two of the authors, and 34 by manual curation of seeds. These sum to 2{,}109; the remaining three candidates are the expert recommendations added after screening had already run, which were coded directly.

The LLM-assisted figure needs qualifying, because it is the part of this pipeline
least able to be re-executed. Those 668 decisions were not one scripted call against
a pinned model with fixed generation settings; they were three passes made by an AI
assistant in interactive sessions---356 in a general screening pass, 180 in an
abstract-scan pass, and 132 in a later refinement pass, distinguished in the released
table. There is therefore no single prompt, model version, or temperature we can
report, and re-running the pipeline would not reproduce the same 668 verdicts. What
we can offer is per-record inspectability: every one of the 668 carries a reason
string recording why it was decided as it was, and every excluded record is retained
with that reason. A reader can audit any individual decision; a reader cannot rerun the pass. Every excluded
record is retained with the rule that excluded it, so the filter can be audited rather
than only described.

\section{Verification and Screening Details}\label{app:validity}

%This appendix carries the supporting detail for Section~\ref{threats-to-validity}. Nothing here is required to follow the results; it is the record a reader would need to challenge one.

\subsection{The two verification routes}
Records were drafted by an LLM from the papers' full text and then verified against the
papers by the authors. For \textbf{80 records} an author checked every graded value against
the source directly. The \textbf{31} merged later from a second extraction pipeline were
additionally screened first against four conflict signals---disagreement with an independent
earlier coding of the same paper, contradiction of the codebook by the record's own other
fields, contradiction of the span recorded for the value, and whether a headline claim
depends on the record---with every flag adjudicated independently by three authors, before
every populated field was checked in turn. The \texttt{verification} field records which
route each record took.

That second pass also yields a value-level error rate. Across \textbf{372 field-checks} on
those 31 records, \textbf{19 values were wrong (5.1\%)}, touching 19 of the 31 records.
Every one of the 19 was the same field: the pipeline had defaulted \emph{provenance} to
\emph{mixed}, where the papers' descriptions put 15 at \emph{human} and 4 at
\emph{AI-generated}. A systematic default in one field is the most tractable thing such a
check can find, and it is now corrected corpus-wide. Provenance is not one of the four
columns of the quality matrix, though it does enter the realism decomposition of
Section~\ref{realism}, so the correction is reflected there.

The risk AI drafting creates is specific: a fluent but unsupported value. Two controls
address it, and both pipelines enforce them. The first is that every rubric-graded value
must carry a verbatim span. Measured against the released corpus, that holds without
exception: all 332 such values across the 111 records are span-backed, and the 31 records
verified by triage contribute 248 of the corpus's 729 spans in total. Two classes of value
are outside the rule. Absence values assert that a paper does not report something, which no span can evidence; Section~\ref{construct-validity} describes the targeted scans used in their place.
Granularity, scale, and language coverage are read off the paper rather than graded against
a rubric. The second control is mechanical: numeric claims were checked against the source,
with 231 scale figures and 155 diversity figures confirmed present in the paper that was
coded.

That check earned its place. It found a record whose scale and label had been taken from a
sentence in the paper's related work describing a \emph{different} dataset. Verification
changed at least one of the 13 coded fields in 31 of the 68 records that existed at the time
of the first reading round; the remaining 12 direct-route records were added afterwards---from
the seed and advisor passes, the audit re-inclusions, and the two Tier-1 candidates admitted
after the re-screen---and were read on the same protocol. That is a record-level rate and
should not be read as a per-value error rate: those 68 records carry 884 value slots, and
because we did not log changes field by field, 31 of 68 bounds the per-value rate from above
rather than measuring it. Verification also exposed two systematic error patterns worth
naming: following a paper's self-description (``real-world dataset'') instead of our own
rubric, and failing to apply our own rule for studies spanning many datasets. Both were
corrected corpus-wide.

\subsection{Agreement, and what triage changed}
Across the 33 triage flags, 94 of the 99 individual reviewer verdicts upheld the coded value (94.9\%), and mean pairwise agreement was \textbf{96\%} (Nong/Du 93.9\%,
Nong/Xu 97.0\%, Du/Xu 97.0\%); the corresponding Cohen's $\kappa$ values are 0.47, 0.65, and
0.65. At the flag level the coded value was retained for 32 of the 33 flags: thirty were
upheld unanimously, one was rejected unanimously, and two split two-to-one in favour of the
coded value.

The triage produced one correction: a T1 dataset moved from \emph{injected/generated} to
\emph{synthetic}, its samples being model-generated from question text rather than real code
carrying an inserted flaw. All three reviewers rejected the coded value; two proposed
\emph{synthetic} and one \emph{mined/semi-real}, and the codebook settles the split, since
mined/semi-real is defined as material collected from real projects or CVEs, which generated
code is not. Separately, and before the triage ran, the merge itself corrected one T1 dataset
from \emph{real-world} to \emph{mined/semi-real}, because it labels CVE-derived functions from
patches, which our codebook defines as mined; the pipeline had followed the paper's own
``authentic CVEs'' phrasing instead of our rubric, the same failure mode first-round
verification found. The surviving T1 real-world claim rests on a single record, a
smart-contract dataset whose samples are whole deployed contracts, and all three reviewers
upheld that value on the ground that the deployed unit and the coded sample coincide.

The 80 records verified field by field were each checked by one author, so for those we have
auditability rather than a measured agreement rate. A record can also be wrong in a way no
screen detects---if both codings share a mistake, or if a span was drawn from the right paper
but the wrong dataset inside it. Twenty-five records carry a confidence below \emph{high}:
nine of the 80 verified field by field (eight medium, one low) and sixteen of the 31 verified
by triage. In almost every case the flag marks a value the paper does not state, rather than
one we could not read; the single low-confidence record is one whose dataset-construction
description is too thin to code safely, and no quantitative claim in this study rests on it.

\subsection{Tagger accuracy by type and period}
The corpus-wide artifact types come from a keyword tagger, and the 111 hand-coded papers let
us measure it: 92 of them carry a usable automatic tag, and against the manual type the
tagger agrees on \textbf{70.7\%} of them (macro-$F_1$ 0.64). Accuracy is uneven by type.
Patches are recovered well (recall 1.00, precision 0.80) and code samples adequately
(0.79 / 0.73), but executable artifacts are recovered only 55\% of the time---though what the
tagger does call T4 is almost always right (precision 0.92)---and reasoning data is recovered
12\% of the time, which is why the corpus-wide reasoning re-screen found six datasets the
tags had missed.

The error is also \emph{not stationary}, which matters more than its size. Split by period,
the tagger recovers 0 of the 3 true T4 papers published through 2023, then 6 of 9 in 2024--25
and 6 of 10 in 2026. The jump is between the earliest band and the rest rather than a steady
climb, but it is enough to disqualify a trend measured from a 2023 baseline.

\subsection{The re-screen of excluded records}
Our corpus-wide selection originally applied each artifact-type filter \emph{within} the set
of papers the automatic tagger had already assigned to that type, so a paper the tagger typed
wrongly could never be reconsidered for another. Re-running the reasoning filter across every
included paper instead surfaced six further reasoning-data contributions---all typed T1, T2 or
T3 by the tagger---and took that row from one dataset to seven; the later T1c pass added the
eighth (Section~\ref{res:t5}). The trace row was re-screened the same way and surfaced two
further releases, both of which we then deep-coded.

We then re-screened all 808 excluded records. We split them into three disjoint sets of
roughly 269 and assigned one author to each, so every record received one fresh judgement
from a reader who did not make the original call and who worked from title and abstract. We
withheld the original exclusion reason deliberately, since it is the strongest anchor
available to a second reader and the point was an independent judgement. Because the sets do
not overlap we report no inter-rater statistic, and no per-record adjudication was possible;
consistency rests instead on a written rule applied by all three, and on a second pass that
corrected that rule where it proved too loose. This returned \textbf{217 papers}, taking the
corpus from 1{,}298 to 1{,}515. The re-inclusions are not spread evenly: 155 carry executable
artifacts and 54 carry traces, so the original screen missed executable and observational work
disproportionately. Mid-exercise we found the eligibility rule underspecified for attack and
defence papers, which a literal reading admitted wholesale; we tightened it to require an
artifact a third party could reuse---a benchmark, a generator, a released harness, or recorded
measurements---rather than an attack demonstrated only in the paper's own results, and
re-decided every affected judgement against it. Two of the three reviewers then arrived at
identical inclusion rates and the third within five points.

We then reapplied the Tier-1 anchor criteria to all 1{,}522 included papers. Under T1a and
T1b none of the 217 qualifies: not one carries a recorded citation, against 43 of the 63
anchors whose OpenAlex metadata resolves, 85\% were published in 2025--26, and 7 appear in an
indexed journal or top-tier venue. Under T1c a deliberately broad keyword screen flagged ten
for reading; nine release agent-security benchmarks or measurements rather than findings about
vulnerability data, and one qualifies~\cite{p486}. Under T1d, four apparent primary trace
corpora sit outside the anchor set (Section~\ref{res:t6}). We read all five in full against
the same procedure every other anchor received: two qualified and are now deep-coded, the
post-training study under T1c~\cite{p486} and one trace corpus under T1d~\cite{p14}. Three did
not---their releases are benchmarks and simulation venues whose runs emit traces, which our
tie-break rule (Section~\ref{sec:aspect-a}) codes as evaluation records rather than primary
trace artifacts. The anchor set is therefore 111 records rather than 109.

\subsection{Estimating residual screening recall}
The re-screen cannot measure its own recall, since its three sets are disjoint and no record
was read twice. A second reader, never the one who first excluded the record, re-screened 122
of the 591 records that remained excluded, allocated across all 41 exclusion categories in
proportion to category size with a floor of two, so that small categories were represented at
all. \textbf{Seven proved eligible and are restored}, which is why the corpus holds 1{,}522
papers rather than 1{,}515.

Three quantities follow and they are distinct. The raw rate among reviewed exclusions is 7 of
122. Because the floor of two over-represents small categories, the size-weighted estimate of
the residual false-exclusion rate is \textbf{4.5\% (95\% CI 0.7--8.4\%)}: a stratified
proportion $\hat{p}=\sum_h (N_h/N)\,(k_h/n_h)$ with design-based variance
$\sum_h (N_h/N)^2 (1-n_h/N_h)\,p_h(1-p_h)/(n_h-1)$ and a normal-approximation interval, the
strata being the exclusion categories and the finite-population correction applying because
several categories were sampled heavily. That compares with the 26.9\% the original screen
missed. Applied to the 469 records that remained excluded and were never re-read, it implies
roughly 21 further eligible papers, putting \textbf{screening recall within the retrieved
candidate pool at 98.6\%} (97.5--99.8\%). This measures how much of what our searches returned we correctly kept; it does not measure retrieval recall, and says nothing about work the queries never returned. The corpus also inherits the language concentration it documents, since we searched in English only.

\subsection{How much coder disagreement the conclusions absorb}
For each headline finding we computed the smallest number of records that would have to be
coded differently, all in the direction least favourable to us, before the finding reverses.
The claim that executable artifacts are the only well-populated type meeting both criteria by a majority survives 2 such recodes and reverses on the third: patch datasets would
join it if 3 were regraded real-world and one of those also became verified, since a record
carries both attributes at once, while T4 itself would cease to hold a majority only after 4.
Counted as individual attribute values rather than records the figure is 4, and we report the
record count because that is the unit a coder works in. That the margin is 2 rather than
larger is a consequence of the row sizes the study reports, 24 executable datasets and 14 patch datasets, and not of the coding being close: the two rates differ by 31 points. The leakage finding is less exposed, surviving 8 adverse recodes. Of the graded values in the matrix, 109 are close to mechanical---whether the paper states a release route, whether it names a split strategy---while 180 rest on a graded judgement about authenticity or about what an independent check amounts to; the second group is where an independent coder would most likely diverge. We restrict these threshold
comparisons to types with at least ten datasets for the same reason
Figure~\ref{fig:matrix} hatches thin cells: in a row of three, one recode moves the rate by 33
points, and a claim that turns on it is not a finding.

\section{Released Materials}\label{app:released}

The following are released with the paper: the harvesting scripts containing the
queries above; the screening rules, and the full candidate table
with every disposition, its \texttt{decided\_by} route, and its reason string; the deep-coding schema and
codebook; the 111 coded records with their 729 evidence spans; the three-reviewer
verification worksheet with each adjudication; the five re-screening worksheets as issued and as
returned---the three disjoint sets covering all 808 excluded records, and the two second-pass sheets
that re-decided the attack-and-defence boundary cases against the reuse rule---each carrying the reviewer's verdict and stated reason, together with the second-reader validation sample over the records that remained excluded; and the analysis scripts that produce the tables and figures in this
paper, together with the checks that reconcile the aggregate counts quoted in the text against the
coded corpus.

%% ACM fills these in at acceptance; the template expects them present.
% \received{XX Month 2026}
% \received[revised]{XX Month 2026}
% \received[accepted]{XX Month 2026}

\bibliographystyle{ACM-Reference-Format}
\bibliography{refs,refs_band2}

\end{document}